\documentclass[preprint2]{aastex631}

\usepackage{graphicx}
\usepackage{natbib}
\usepackage{booktabs}
\usepackage{amsmath, physics}

\newcommand{\Cornell}{Cornell Center for Astrophysics and Planetary Science, and Department of Astronomy, Cornell University, Ithaca, NY 14853, USA}
\newcommand{\Columbia}{Department of Astronomy, Columbia University, 550 West 120th Street, New York, NY 10027, USA}
\newcommand{\KZA}{University of Malta, Institute of Space Sciences and Astronomy, Malta}
\newcommand{\NIJ}{Department of Astrophysics/IMAPP, Radboud University, Nijmegen, Netherlands}
\newcommand{\SETI}{SETI Institute, Mountain View, CA, USA}
\newcommand{\UCB}{Department of Astronomy,  University of California Berkeley, Berkeley CA 94720, USA}
\newcommand{\Curtin}{International Centre for Radio Astronomy Research, Curtin Institute of Radio Astronomy, Curtin University, Perth, WA 6845, Australia}

\shorttitle{The Galactic Center Magnetar at 4--8~GHz}
\shortauthors{Suresh et al.}
\begin{document}

\title{4--8~GHz Spectro-temporal Emission from the Galactic Center Magnetar PSR~J1745$-$2900}

\correspondingauthor{Akshay Suresh}
\email{as3655@cornell.edu}

\author[0000-0002-5389-7806]{Akshay Suresh}
\affiliation{\Cornell}

\author[0000-0002-4049-1882]{James M. Cordes}
\affiliation{\Cornell}

\author[0000-0002-2878-1502]{Shami Chatterjee}
\affiliation{\Cornell}

\author[0000-0002-8604-106X]{Vishal Gajjar}
\affiliation{\UCB}

\author[0000-0002-6341-4548]{Karen I. Perez}
\affiliation{\Columbia}

\author[0000-0003-2828-7720]{Andrew P. V. Siemion}
\affiliation{\UCB}
\affiliation{\NIJ}
\affiliation{\SETI}
\affiliation{\KZA}

\author[0000-0003-2783-1608]{Danny C. Price}
\affiliation{\UCB}
\affiliation{\Curtin}
%% Keep to less than 220 words.
\begin{abstract}
Radio magnetars are exotic sources noted for their diverse spectro-temporal phenomenology and pulse profile variations over weeks to months. Unusual for radio magnetars, the Galactic Center (GC) magnetar PSR~J1745$-$2900 has been continually active since its discovery in 2013. We monitored the GC magnetar at 4--8~GHz for 6~hours in August--September 2019 using the Robert C.~Byrd Green Bank Telescope. During our observations, the GC magnetar emitted a flat fluence spectrum over 5--8~GHz to within $2\sigma$ uncertainty. From our data, we estimate a 6.4~GHz period-averaged flux density, $\overline{S}_{6.4} \approx (240 \pm 5)~\mu$Jy. Tracking the temporal evolution of $\overline{S}_{6.4}$, we infer a gradual weakening of GC magnetar activity during 2016--2019 relative to that between 2013--2015.5.  Typical single pulses detected in our study reveal marginally resolved sub-pulses with opposing spectral indices, a feature characteristic of radio magnetars but unseen in rotation-powered pulsars. However, unlike in fast radio bursts, these sub-pulses exhibit no perceptible radio frequency drifts. Throughout our observing span, $\simeq 5$~ms scattered pulses significantly jitter within two stable emission components of widths, 220~ms and 140~ms, respectively, in the average pulse profile.
\end{abstract}

% SECTION 1: INTRODUCTION
\section{Introduction} \label{sec:intro}
Magnetars are young rotating neutron stars that emit intense electromagnetic radiation (see reviews by \citealt{Kaspi2017} and \citealt{Esposito2021}) powered by the decay of their enormous internal magnetic fields ($B \sim 10^{13}$--$10^{15}$~G; \citealt{Duncan1992, Thompson1995, Thompson1996}). Prominent features of transient magnetar emission include millisecond bursts and month-long flares, particularly at X-ray and soft $\gamma$-ray wavelengths. To date, 25 Galactic magnetars \citep{Olausen2014}\footnote{McGill Online Magnetar Catalog: \url{http://www.physics.mcgill.ca/~pulsar/magnetar/main.html}} have been confirmed, of which only five have been seen to display pulsed radio emission \citep{Camilo2006,Camilo2007a,Levin2010,Eatough2013,Shannon2013,Esposito2020,Lower2020}. \\

The Galactic Center magnetar PSR~J1745$-$2900 has been continuously active at radio frequencies ($\nu$) since its discovery in 2013 \citep{Eatough2013}. Like other radio-loud magnetars, its mean flux density ($S_{\nu}$), pulse-averaged profile, and emission spectral index ($\alpha = \text{d}(\log S_{\nu})/\text{d}(\log \nu)$) exhibit substantial variability over weeks to months \citep{Lynch2015,Torne2015,Torne2017,Pearlman2018,Wharton2019}. In addition, its average pulse profile typically contains multiple emission components that evolve significantly with radio frequency. \\

Single pulses from the GC magnetar often comprise narrow ``spiky'' sub-pulses \citep{Yan2015}, evoking comparisons with rotating radio transients (RRATs: \citealt{McLaughlin2006}), pulsar giant pulses (GPs: \citealt{Johnston2004}), and fast radio bursts (FRBs: \citealt{Cordes2019,Petroff2019,Chatterjee2021}). However, unlike RRATs and GPs, magnetar single pulses can manifest with diverse morphology between rotations \citep{Pearlman2018}. Additionally, they show no evidence of the ``sad trombone'' structure (negative radio frequency drift with increasing arrival time; \citealt{Hessels2019,Fonseca2020}) that is characteristic of some repeating FRBs. \\ 

Dissimilar to rotation-powered pulsars ($S_{\nu} \propto \nu^{-1.4 \pm 1.0} $; \citealt{Bates2013}) radio magnetars usually exhibit flat or inverted spectra \citep{Levin2012,Torne2015,Torne2017,Dai2019}. Nonetheless, \citet{Pearlman2018} obtained an emission spectral index, $\alpha = -2.08 \pm 0.04$, for the GC magnetar on 2015 July~30. More recently, \citet{Lower2021} observed emission mode switching in the radio magnetar Swift~J1818.0$-$1607 over a 5-month window. A pulsar-like emission spectrum with $\alpha = -1.7^{+0.2}_{-0.3}$ on 2020 May~8 flattened to $\alpha = -0.5 \pm 0.1$ by 2020 October~6. Continued monitoring of radio magnetars is essential for identifying potential links between rotation-powered pulsars and magnetars. \\
\vspace{-10mm} % Remove space before Table 1.

%%%% TABLE 1
\begin{deluxetable*}{CccCCCcCC}
\tablecaption{Log of 4.4--7.8~GHz observations analyzed in our GC magnetar study. \label{tab1}}
\tablewidth{0pt}
\tablehead{
\colhead{Epoch} & \colhead{Calibrator} & \colhead{Test pulsar} & \colhead{Scan} & \colhead{Start MJD} & \colhead{Duration} & \colhead{$\theta_{\rm{GC}}$\tablenotemark{a}} & \colhead{$N_{\rm pulses}$\tablenotemark{b}} & \colhead{$f_P$\tablenotemark{c}} \\
\colhead{(number)} & \colhead{} & \colhead{} & \colhead{(number)} & \colhead{(UTC)} & \colhead{(min.)} & \colhead{(deg.)} & \colhead{(number)} & \colhead{}
}
\startdata
1 & 3C~286 & J2022$+$5154 & 1.1 & 58705.175 & 60 & 8--15 & 663 & 0.32 \\
\hline 
2 & \nodata & \nodata & 2.1 & 58734.958 & 30 & 21--22~~ & 222 & 0.24 \\
& & & 2.2 & 58734.979 & 30 & 22--23~~ & 230 & 0.24 \\
\hline
3 & 3C~286 & J2022$+$5154 & 3.1 & 58737.962 & 30 & 22--23~~ & 380 & 0.38 \\
& & & 3.2 & 58737.983 & 30 & 22--23~~ & 367 & 0.37 \\
& & & 3.3 & 58738.004 & 30 & 21--22~~ & 368 & 0.36 \\
& & & 3.4 & 58738.025 & 30 & 20--21~~ & 380 & 0.37 \\
& & & 3.5 & 58738.046 & 30 & 18--20~~ & 346 & 0.36 \\
& & & 3.6 & 58738.067 & 30 & 15--18~~ & 353 & 0.35 \\
& & & 3.7 & 58738.088 & 30 & 11--15~~ & 337 & 0.33 \\
& & & 3.8 & 58738.109 & 30 & ~7--11 & 345 & 0.34
\enddata
\tablenotemark{a}{Elevation range spanned by the GC during scan}
\tablenotetext{b}{Number of GC magnetar pulses detected with matched filtering S/N, ${\rm (S/N)}_{\rm mf} \geq 8$}
\tablenotetext{c}{Fraction of GC magnetar rotations with ${\rm (S/N)}_{\rm mf} \geq 8$ single pulse detections}
\vspace{-8mm}
\end{deluxetable*}
%%%%%%%%%%%%%%%%%%%%%%%%%%%%%%%%%%

Located $\simeq 2\farcs4$ from Sgr~A* \citep{Kennea2013}, observations of the GC magnetar enable probes of the turbulent, central interstellar medium (ISM) of our Galaxy. Using 1--19~GHz observations of the GC magnetar, \citet{Spitler2014} derived the pulse-broadening time scale,
%%%% EQUATION 1
\begin{align}\label{eqn1}
\tau_{\rm sc} (\nu) \simeq 1.3 \pm 0.2~{\rm s}~\left( \frac{\nu}{1~{\rm GHz}} \right)^{-3.8 \pm 0.2},  
\end{align}
%%%%%%%%%%%%%%%%%%%%%%%%
which is nearly three orders of magnitude smaller than that predicted by the NE2001 Galactic electron density model \citep{Cordes2002}. Tracking the rotation measures ($|\rm{RM}| \simeq 6.4$--$6.6 \times 10^4$~rad~m$^{-2}$) and dispersion measures (DM $\simeq$ 1760--1780~pc~cm$^{-3}$) of single pulses from the GC magnetar, \citet{Desvignes2018} noted a $5\%$ fractional $|\rm{RM}|$ decline between 2013--2017 with minimal DM variations. An analogous $|\rm{RM}|$ decrease with minute DM fluctuations has lately been observed for FRB~121102 \citep{Hilmarsson2021}, the only other known source with $|\rm{RM}| \sim 10^5$~rad~m$^{-2}$ \citep{Michilli2018} comparable to that of the GC magnetar. Regular observing of the GC magnetar is necessary to facilitate further comparisons with FRB~121102, and better understand magnetar radio emission. \\

Here, we present results from a 4--8 GHz study of the GC magnetar using data from the Robert C.~Byrd Green Bank Telescope (GBT). These data were acquired as part of the Breakthrough Listen Galactic Center search for intelligent life \citep{Gajjar2021}. Section~\ref{sec:obs} describes our observations and data pre-processing. We detail our single pulse and periodicity analyses of the GC magnetar in Sections~\ref{sec:singlepulse} and \ref{sec:periodicity} respectively. Finally, we summarize our key findings, and discuss their significance in Section~\ref{sec:disc}. 

% SECTION 2: OBSERVATIONS
\section{Observations} \label{sec:obs}
Table~\ref{tab1} presents an overview of our GBT observations organized as three epochs corresponding to the Modified Julian Dates (MJDs) 58705 (2019 August~10), 58735 (2019 September~9), and 58738 (2019 September~12). Each epoch consisted of one or more GC scans of length at least 30~minutes. To verify our data integrity, we recorded 5-minute scans on the test pulsar J2022+5154 (B2021$+$51) during epochs~1 and~3. In addition, we measured our sensitivity at these epochs via position switching on the flux density calibrator 3C~286. For position switching, we supplemented every 2-minute scan on 3C~286 with a 2-minute off-source pointing directed $1\degr$ away from the calibrator sky position. \\

All observations utilized the C-band receiver and the Breakthrough Listen Digital Backend \citep{MacMahon2018,Lebofsky2019}. For details of our backend setup and data product generation, refer to \citet{Gajjar2021}. Here, we work with GC data having $\approx 43.69~\mu$s time resolution and $\approx$~91.67~kHz spectral resolution. These total intensity data contain 53248~channels spanning 3.56--8.44~GHz, thus providing 4.88~GHz of bandwidth overlapping the 3.9--8.0~GHz instantaneous coverage of the C-band receiver.

\subsection{Data Pre-processing}\label{sec:preprocessing}
We searched our data for radio frequency interference (RFI) using the {\tt rfifind} routine of the pulsar software suite {\tt PRESTO} \citep{PRESTO}. The {\tt rfifind} task (see Section~3.4.2 of \citealt{Lazarus2015} for details) computes statistics of time-radio frequency data (dynamic spectra), and outputs a mask listing the set of channels and time blocks to be flagged. Running {\tt rfifind} with an integration time of 1~s, we detected persistent narrow-band RFI between 4.24--4.39,  4.90--4.95, and 6.90--7.10~GHz. Clipping bandpass edges and applying our {\tt rfifind} mask, the usable frequency band in our data extends between 4.4--7.8~GHz. Broadband, short-duration ($\lesssim 1$~s) RFI affected $< 1\%$ of time integrations. \\

After RFI excision, we dedispersed our test pulsar data at trial DMs between 0--50~pc~cm$^{-3}$ (both limits included),  with a grid spacing of 0.5~pc~cm$^{-3}$. We then executed matched filtering searches for single pulses, and folding searches for periodic pulsations in the resulting dedispersed time series. In doing so, we recovered signal-to-noise-maximizing DMs, rotational periods, and average pulse profiles consistent with known properties of PSR~J2022$+$5154 \citep{Manchester2001}\footnote{\url{https://www.atnf.csiro.au/research/pulsar/psrcat}}.

\subsection{Flux Density Calibration}\label{sec:calibration}
Following standard single dish calibration techniques \citep{O'Neil2002}, we modeled the net system temperature ($T_{\rm sys}^{\rm GC}$) towards the GC as follows.
%%%%%%% EQUATION 2
\begin{align}\label{eqn2}
T_{\rm sys}^{\rm GC}(\nu,~t) =~& T_{\rm Rg}(\nu) + (T_{\rm GC}(\nu) + T_{\rm CMB})e^{-A(t)\tau_{\nu}} \nonumber \\
&+ T_{\rm atm}(1-e^{-A(t)\tau_{\nu}}).
\end{align}
%%%%%%%%%%%%%%%%%%%%%%%%%%%%%%%%%%%
Here, $T_{\rm GC} (\nu) \simeq 568~{\rm K} \left(\nu/1~{\rm GHz} \right)^{-1.13}$ \citep{Rajwade2017} is the background continuum temperature in the direction of the GC for a single dish radio telescope with the same aperture as the GBT. Further, $T_{\rm CMB} \approx 2.73~{\rm K}$ is the isotropic Cosmic Microwave Background (CMB) temperature \citep{Fixsen2009}, and $T_{\rm atm} \approx 267~{\rm K}$ is the atmospheric temperature\footnote{\url{http://www.greenbankobservatory.org/~rmaddale/WeatherGFS3/tatm.html}} at 6~GHz.  While $\tau_{\nu} \approx 10^{-4} (80 + 1.25e^{\sqrt{(\nu/1~{\rm GHz})}})$ represents the zenith atmospheric opacity\footnote{\url{https://www.gb.nrao.edu/GBT/DA/gbtidl/gbtidl_calibration.pdf}}, $A (t) = 1/\sin \theta_{\rm GC} (t)$ measures the average airmass at the elevation, $\theta_{\rm GC} (t)$, of the GC \\

\begin{figure}[t]
\centering
\includegraphics[width=0.5\textwidth]{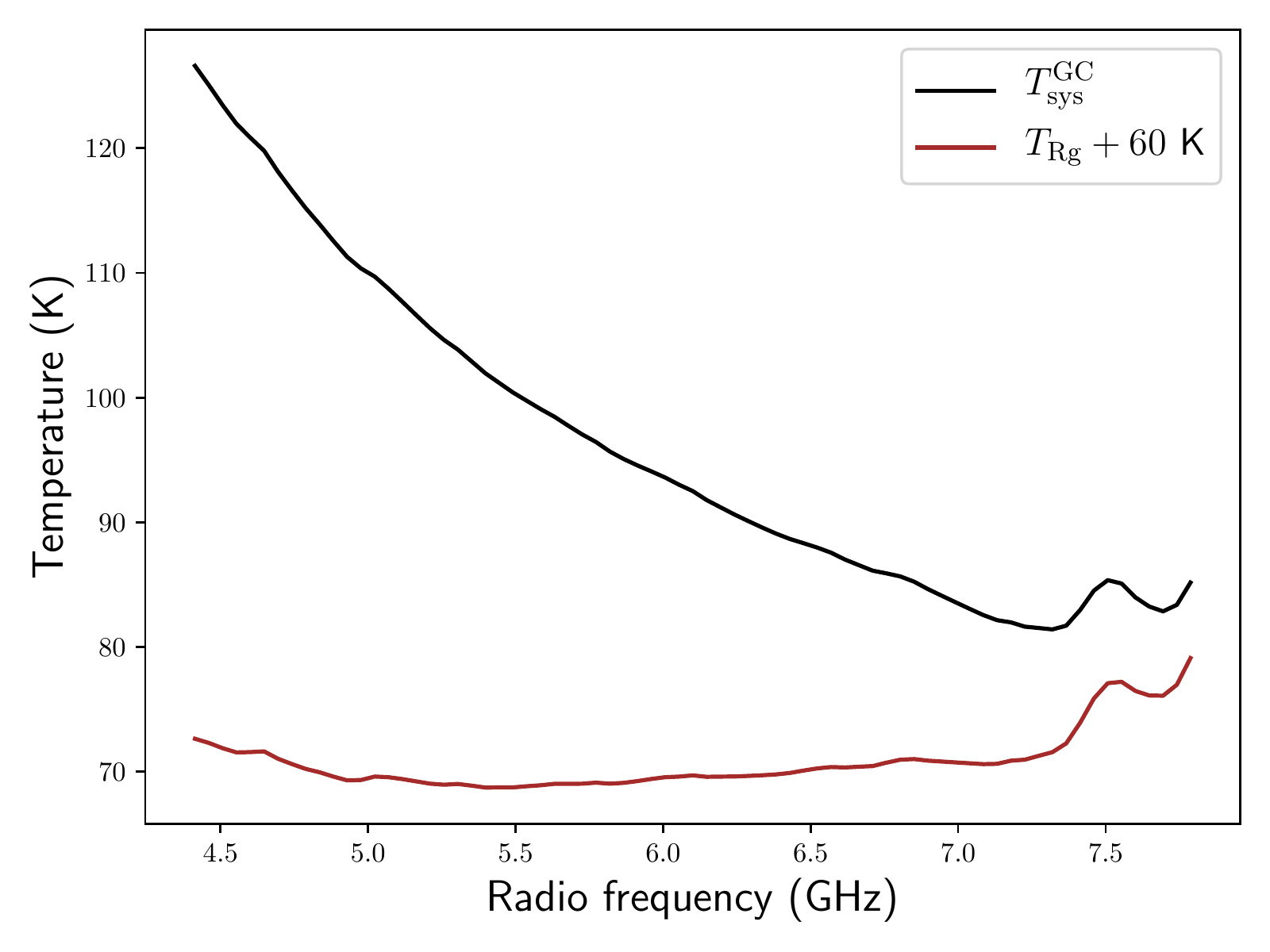}
\caption{Spectra of $T_{\rm Rg}(\nu)$ (brown, translated up by 60~K) and the time-averaged $T_{\rm sys}^{\rm GC}$ (black). While $T_{\rm GC} (\nu)$ in Equation~\ref{eqn2} governs the broadband spectral variation of $T_{\rm sys}^{\rm GC}$, the bump in $T_{\rm sys}^{\rm GC}$ at 7.5~GHz arises from narrowband structure in $T_{\rm Rg}(\nu)$. \label{fig1}}
\end{figure}

The term $T_{\rm Rg}(\nu)$ in Equation~\ref{eqn2} incorporates noise contributions from the receiver, ground pickup, and spillover. We determined $T_{\rm Rg}(\nu)$ (band-averaged value $\approx$ 11~K) through position switching on our flux density calibrator 3C~286. Using Equation~\ref{eqn2}, we then computed $T_{\rm sys}^{\rm GC} (\nu,~t)$, the time-averaged spectrum of which is shown in Figure~\ref{fig1}. \\
%%% FIGURE 8
%%%%%%%%%%%%%%%%%%%%%%%%%%%%%%%%%% 

To alleviate strong, variable ground pickup and spillover at low elevations, we restricted our calibration to scans 3.1--3.6 with $\theta_{\rm GC} \geq 15\degr$ (see Table~\ref{tab1}). For GBT gain, $G \approx 2~{\rm K~Jy}^{-1}$ at 4--8~GHz, we then derived calibrated flux densities, $S_{\nu}(t)$, from bandpass-corrected dynamic spectra, $D_{\nu}(t)$, via
%%%% EQUATION 3
\begin{align}\label{eqn3}
S_{\nu}(t) = \frac{T_{\rm sys}^{\rm GC} (\nu,~t)D_{\nu}(t)}{G}. 
\end{align}
%%%%%%%%%%%%%%%%%%%%%%%%%%%%%%%%%%
Finally, we computed uncertainties on $S_{\nu} (t)$ assuming a $5\%$ error on our flux calibrator spectrum, and $15\%$ errors on $T_{\rm GC}(\nu)$ and $T_{\rm atm}$.

% SECTION 3: SINGLE PULSE STUDY
\section{Single Pulse Study} \label{sec:singlepulse}
Traditional burst search algorithms generally perform matched filtering of dedispersed time series with template filters of various widths. Incorporating our {\tt rfifind} mask, we first eliminated $\rm{DM} = 0$~pc~cm$^{-3}$ signals by subtracting the mean across channels from each time slice in our dynamic spectra. To detect single pulses from the GC magnetar, we then dedispersed our data at trial DMs between 1600-2000~pc~cm$^{-3}$ (both limits included), with DM step size, $\rm \delta DM =$~0.5~pc~cm$^{-3}$. We subsequently block-averaged our dedispersed time series to $\approx$~350~$\mu$s resolution, and passed these through a boxcar matched filtering search. We trialed boxcar filter lengths of 1, 2, 3, 4, 6, 9, 14, 20, and 30 bins, thus covering $\approx$~0.4--10.5~ms burst widths. \\

%%% FIGURE 2
\begin{figure}[t!]
\centering
\includegraphics[width=0.48\textwidth]{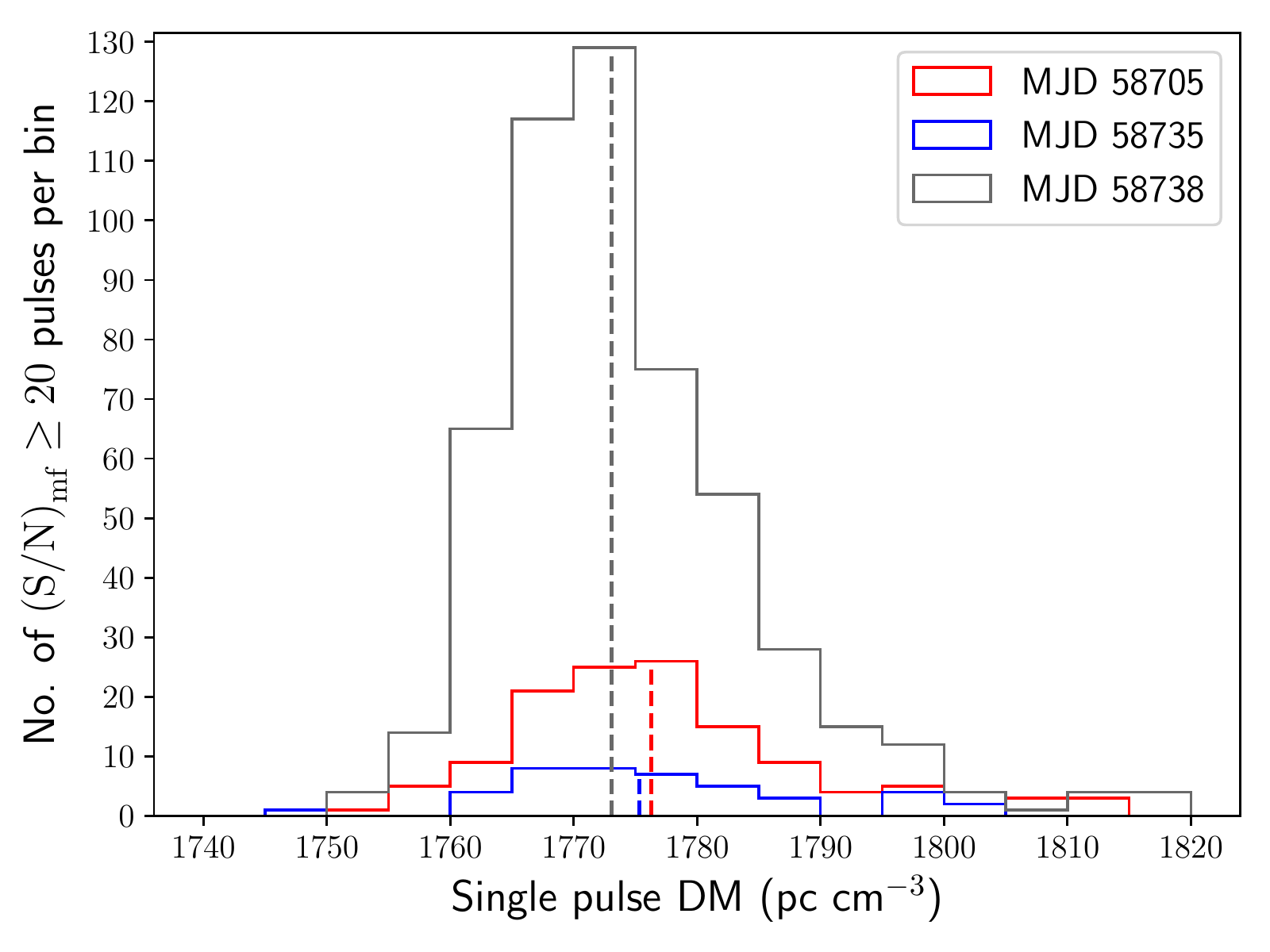}
\caption{Histograms of (S/N)$_{\rm mf}$-maximizing DMs for epochs 1 (red), 2 (blue), and 3 (grey). The vertical dashed lines close to the histogram peaks denote (S/N)$_{\rm mf}^2$-weighted pulse-averaged DMs at their respective epochs. Only single pulses with (S/N)$_{\rm mf} \geq 20$ were considered for DM measurement. \label{fig2}}
\end{figure}
%%%%%%%%%%%%%%%%%%%%%%%%%%%%%%%%%% 
Let (S/N)$_{\rm mf}$ denote the signal-to-ratio of a candidate pulse in the convolution of its dedispersed time series with an optimal boxcar matched filter. Setting (S/N)$_{\rm mf} \geq 8$ as the detection criterion that excludes noisy peaks, Table~\ref{tab1} enumerates single pulse detection counts for our GC scans. We confirmed the astrophysical nature of all (S/N)$_{\rm mf} \geq 8$ candidates through manual visual inspection of their dynamic spectra for continuous $\nu^{-2}$ dispersive sweeps and natural spectro-temporal sub-structure (analogous to that known in pulsars, magnetars, and FRBs). \\

Assuming a barycentric rotation period of $P_0 \approx 3.7686$~s (see Table~\ref{tab2}), we detected single pulses in $f_P \approx$~30--40$\%$ of GC magnetar rotations. But, during epoch~2, this detection fraction dropped to $f_P \approx$~22--23$\%$. In addition, the mean pulse (S/N)$_{\rm mf}$ at epoch~2 was only $\approx 11.3$, while the corresponding numbers for epochs~1 and 3 were about 12.4 and 12.3, respectively. Possible causes for the apparent $f_P$ decline and the comparatively low pulse-averaged (S/N)$_{\rm mf}$ at epoch~2 may be a brief GC magnetar weakening, or an increased $T_{\rm sys}^{\rm GC}$ relative to other epochs. Unfortunately, we cannot distinguish between these scenarios owing to our lack of calibrator observations at epoch~2.
%%%% TABLE 2
\begin{deluxetable}{CClC}
\tablecaption{DM and barycentric period measurements for the GC magnetar. \label{tab2}}
\tablewidth{0.48\textwidth}
\tablehead{
\colhead{Epoch} & \colhead{$\overline{\rm DM}$\tablenotemark{a}}  &  \colhead{$P_0$\tablenotemark{b}} & \colhead{N$_{\rm harm}$\tablenotemark{c}} \\
\colhead{(MJD)} & \colhead{(pc~cm$^{-3}$)} & \colhead{(s)} & \colhead{(number)} 
}
\startdata
58705 & 1776.3 \pm 17.6 & 3.7686(4) & 10\\
58735 & 1775.3 \pm 30.9 & 3.769(1) & \phn 8  \\
58738 & 1773.1 \pm 9.1 \phn & 3.7686(8) & 10 \\
\enddata
\tablenotetext{a}{Average DM computed according to Equation~\ref{eqn4}}
\tablenotetext{b}{Rotation period derived from the highest harmonic of $f_0 = 1/P_0$ seen in power spectrum of barycentric ${\rm DM} =$~1775~pc~cm$^{-3}$ time series. Parenthesized numbers reflect uncertainties on last significant digit of $P_0$.}
\tablenotetext{c}{Number of harmonics of $f_0$ seen in power spectrum of ${\rm DM} =$~1775~pc~cm$^{-3}$ time series}
\vspace{-8mm}
\end{deluxetable}
%%%%%%%%%%%%%%%%%%%%%%%%%%%%%%%%%%

\subsection{DM Measurement}\label{sec:DM_GCmag}
Accurate DM determination requires broadband pulse detection with high S/N. Considering bursts with (S/N)$_{\rm mf} \geq 20$, Figure~\ref{fig2} shows histograms of (S/N)$_{\rm mf}$-maximizing DMs for all observing epochs. For a given epoch with $N$ burst detections, let ${\rm DM}_i$ represent the (S/N)$_{\rm mf}$-maximizing DM of pulse $i$. We then construct the (S/N)$_{\rm mf}^2$-weighted pulse-averaged DM for each epoch as
%%%% EQUATION 4
\begin{align}\label{eqn4}
\overline{\rm{DM}} &= \frac{\sum\limits_{i=1}^{N} {\rm (S/N)}_{{\rm mf},i}^2~{\rm DM}_{i} }{\sum\limits_{i=1}^N {\rm (S/N)}_{{\rm mf},i}^2}.    
\end{align}
%%%%%%%%%%%%%%%%%%%%%%%%%%%%%%%%%%
The DM uncertainty associated with pulse detection across observing bandwidth $B$ at center frequency $\nu_c$ is \citep{Cordes2003}
%%%% EQUATION 4
\begin{align}\label{eqn5}
\Delta {\rm DM} & \simeq 506~{\rm pc~cm}^{-3} \left( \frac{W_{\rm eff,ms}~ \nu_{c,\rm{GHz}}^3}{B_{\rm MHz}} \right).   
\end{align}
%%%%%%%%%%%%%%%%%%%%%%%%%%%%%%%%%%
Here, $W_{\rm eff} \simeq 5.1$~ms is the effective pulse width (see Section~\ref{sec:acf}) in a dedispersed time series. For our observations, $B \approx 3.4$~GHz and $\nu_c \approx 6.1$~GHz, which together imply $\Delta {\rm DM} \simeq 171.3$~pc~cm$^{-3}$. Applying standard error propagation rules to Equation~\ref{eqn4} with $\Delta {\rm DM}_i = \Delta {\rm DM}$, we derived uncertainties on $\overline{\rm DM}$. \\

%%% FIGURE 3
\begin{figure*}[t!]
\gridline{\fig{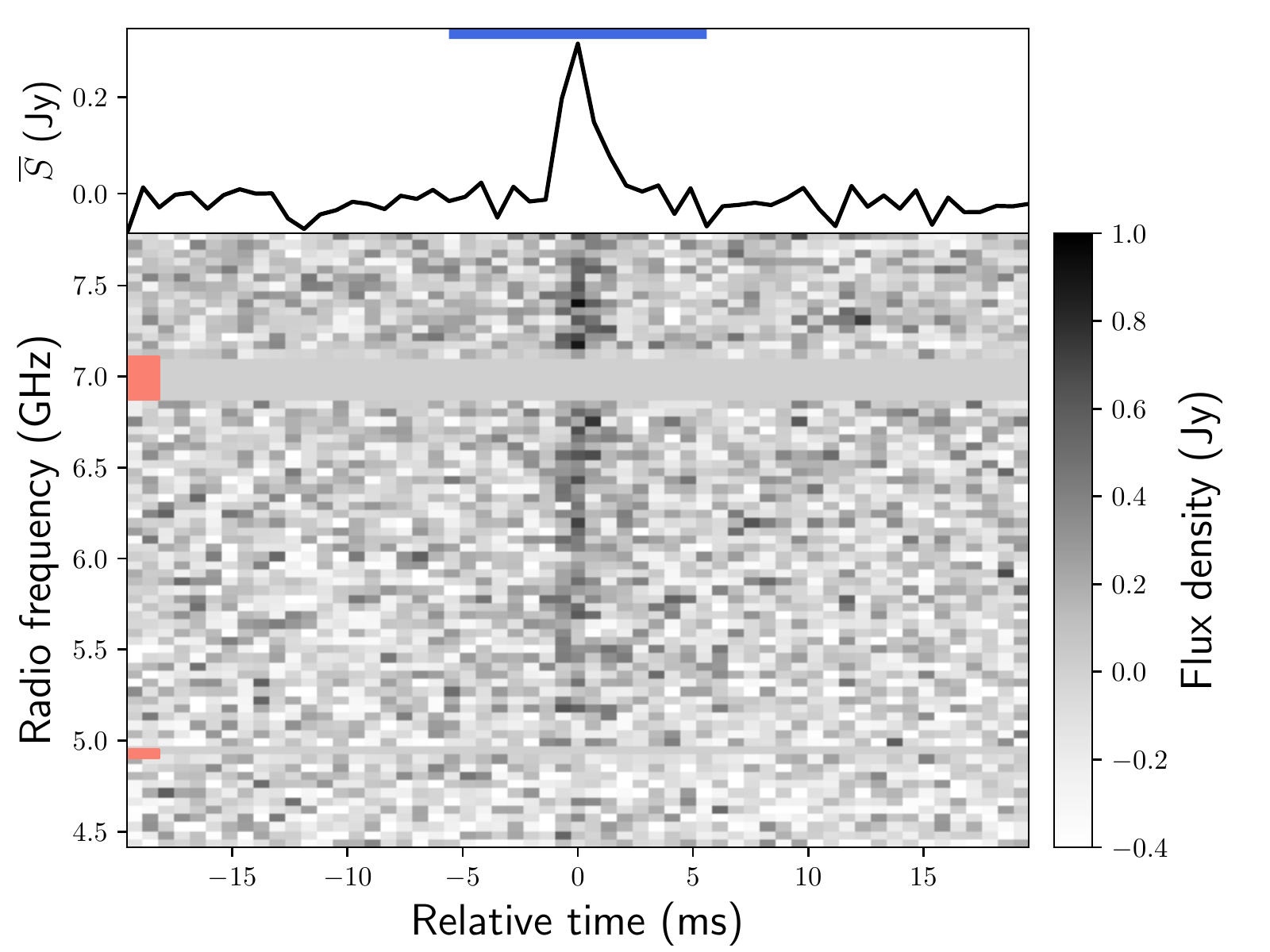}{0.5\textwidth}{(a)}
          \fig{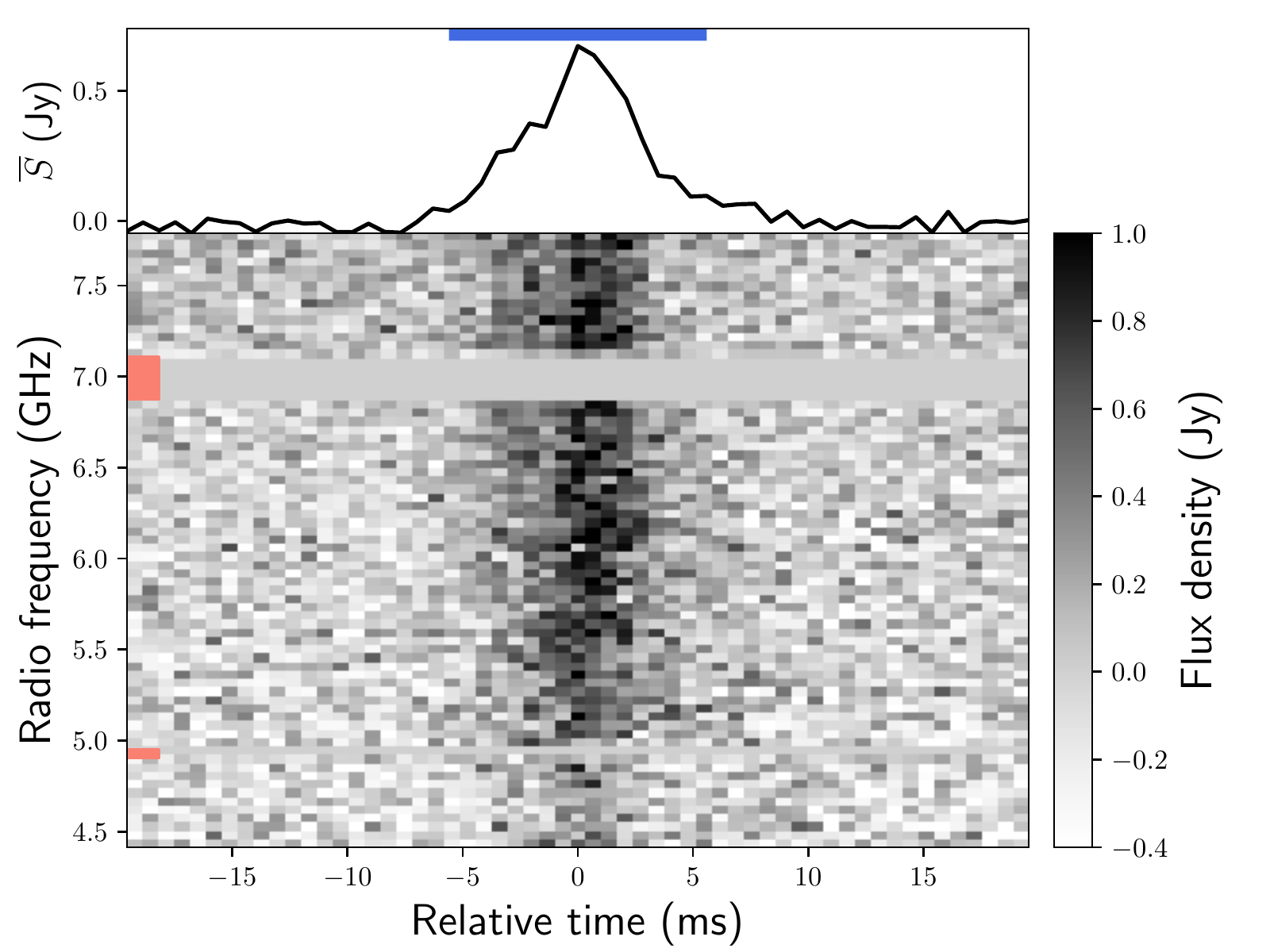}{0.5\textwidth}{(b)} }
\gridline{\fig{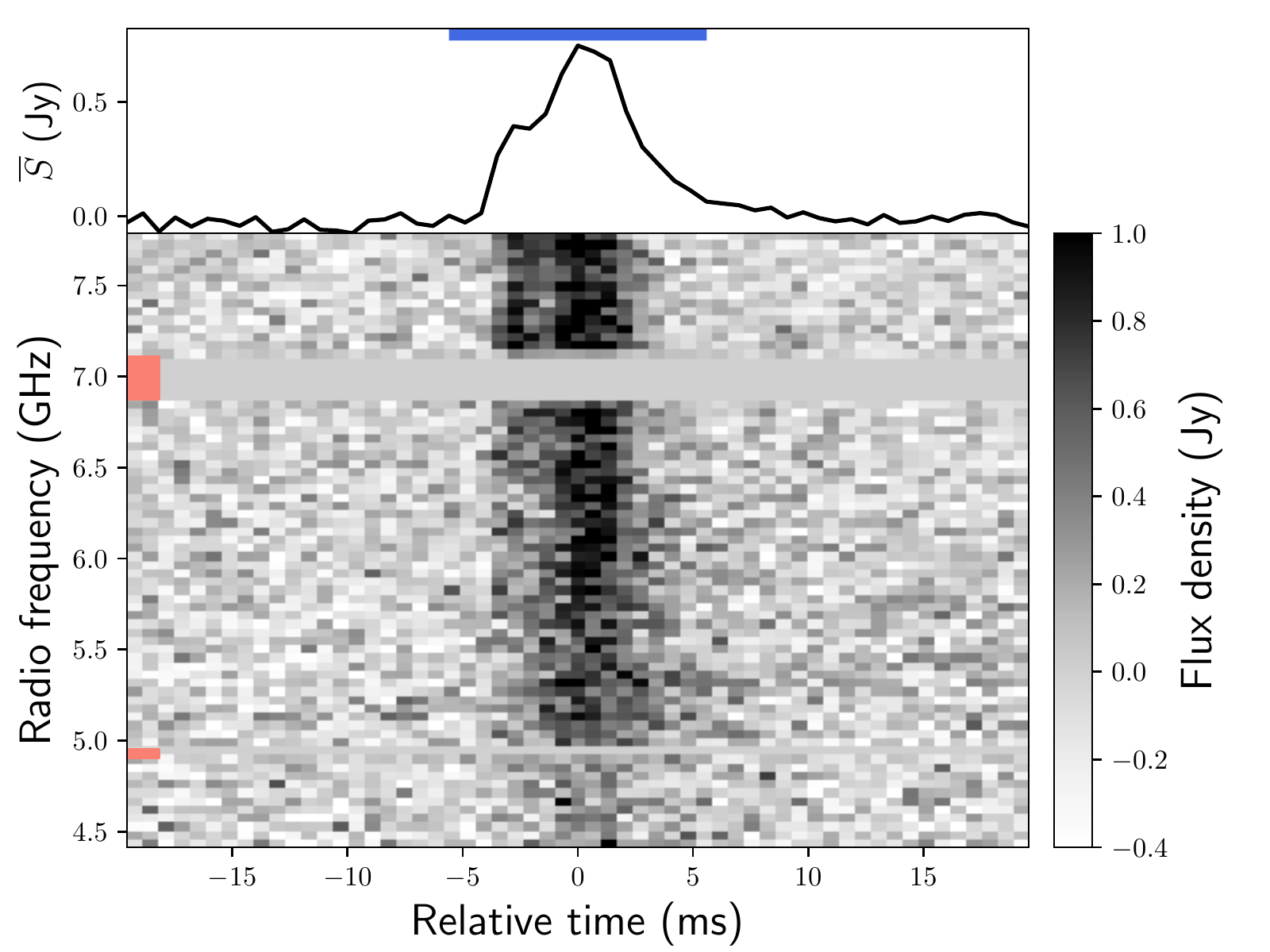}{0.5\textwidth}{(c)}
          \fig{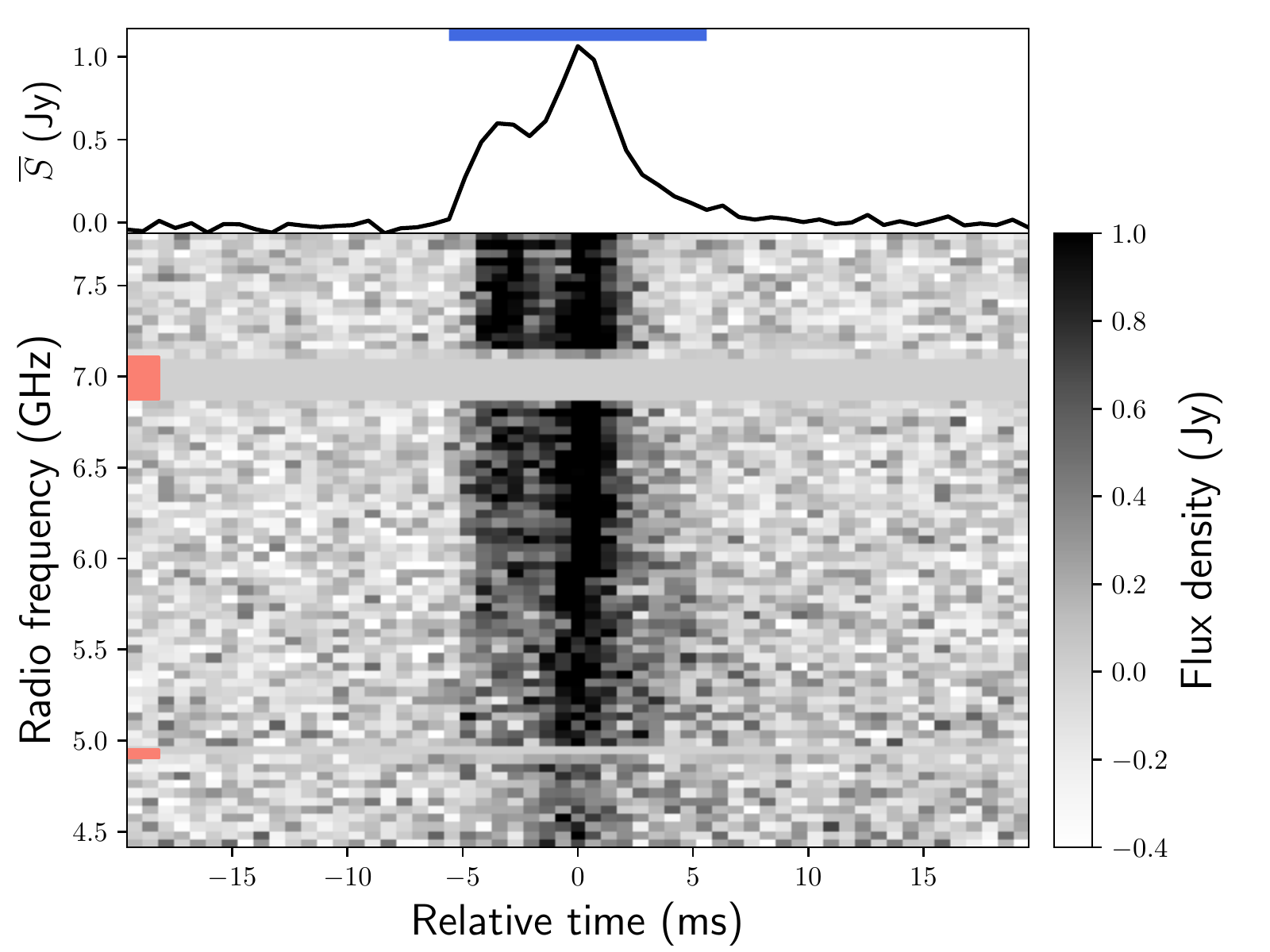}{0.5\textwidth}{(d)} }
\caption{Calibrated, dedispersed dynamic spectra (bottom subplot in each panel), and frequency-averaged time series (top subplot in each panel) of four GC magnetar bursts detected at epoch~3 (MJD~58738). All plotted dynamic spectra have been block-averaged to 47~MHz spectral resolution and 0.7~ms time resolution, yielding an off-pulse standard deviation, $\sigma \approx 0.2$~Jy. The grayscale axes in all panels are set to range between $-0.4$~Jy ($-2\sigma$) and 1~Jy (5$\sigma$). Orange horizontal bars at left edges of all dynamic spectra label flagged channels. Blue horizontal bars in top subplots of all panels indicate the time window ($\pm 6$~ms around single pulse maximum) chosen for computing burst properties. Plausible reasons for the observed abrupt emission decline below 5~GHz may be an instrumental bandpass issue between 4.4--5~GHz, unknown complexities in $T_{\rm GC} (\nu)$, or a spectral turnover intrinsic to the GC magnetar during our observations. \label{fig3}}
\end{figure*}
%%%%%%%%%%%%%%%%%%%%%%%%%%%%%%%%%%

Table~\ref{tab2} lists $\overline{\rm DM}$ measurements for different epochs along with their respective uncertainties. Within $1\sigma$ limits, we find no evidence of inter-epoch $\overline{\rm DM}$ evolution. Hence, we fix ${\rm DM} = 1775$~pc~cm$^{-3}$ for the GC magnetar throughout the remainder of our study.

\subsection{GC Magnetar Burst Characterization}\label{sec:burst_ch}
Figures~\ref{fig3}(a)--\ref{fig3}(d) present calibrated, dedispersed dynamic spectra of four sample GC magnetar pulses from epoch~3. While all bursts contain scatter-broadened tails at late times (relative to pulse maxima), several pulses exhibit fine spectro-temporal sub-structure at early times. For example, as evident in Figures~\ref{fig3}(c) and \ref{fig3}(d), broad pulse envelopes occasionally reveal narrow overlapping sub-pulses with no perceptible radio frequency drifts. \\

Such sub-pulses, akin to all GC magnetar pulses detected in our data (also see Figure~15 of \citealt{Gajjar2021}), show an apparent sharp emission decline below 5~GHz. Furthermore, resolved, leading sub-pulses within wide burst profiles (see Figure~\ref{fig3}(d)) often manifest steeper spectra than their trailing companions. As temporal overlaps between sub-pulses can vary dramatically from one burst to the next, accurate statistical characterization of sub-pulse emission is difficult. Hence, for studying burst properties, we treat broad pulse envelopes as solitary bursts with chromatic intrinsic widths. Sections~\ref{sec:acf} and \ref{sec:skewness} discuss our investigation of GC magnetar pulse widths and asymmetry, respectively. In Section~\ref{sec:fluence}  we study the in-band spectrum of GC magnetar emission at epoch~3. 

\subsubsection{Single Pulse Widths}\label{sec:acf}
Consider a radio burst of intrinsic width $W_{\rm int}(\nu)$ in a dynamic spectrum with channel bandwidth, $\Delta \nu \approx 91.67$~kHz. Its effective width in a dedispersed time series with sample interval, $t_{\rm samp} \approx 350$~$\mu$s, is then \citep{Cordes2003}
%%% EQUATION 6
\begin{align}\label{eqn6}
W_{\rm eff} = \left(W_{\rm int}^2 + t_{\rm samp}^2 + \tau_{\rm sc}^2 + t_{R}^2 + t_{\rm chan}^2 + t_{\rm BW}^2 \right)^{1/2}.    
\end{align}
%%%%%%%%%%%%%%%%%%%%%%%%%%%%%%%%%% 
Here, $t_R \sim (\Delta \nu)^{-1} \approx 11$~$\mu$s is the receiver filter response time. The terms $t_{\rm chan}$ and $t_{\rm BW}$ represent, respectively, the intrachannel dispersive smearing and the residual broadband dispersive delay, which are given by
%%% EQUATION 7
\begin{align}\label{eqn7}
t_{\rm chan} \simeq 8.3~\mu{\rm s} \left(\frac{{\rm DM}_{\rm pc~cm^{-3}} \ \Delta \nu_{\rm MHz}}{\nu_{\rm GHz}^3} \right),
\end{align}  
%%%%%%%%%%%%%%%%%%%%%%%%%%%%%%%%%% 
%%% EQUATION 8
\begin{align}\label{eqn8}
t_{\rm BW} \simeq 8.3~\mu{\rm s} \left(\frac{{\rm \delta DM}_{\rm pc~cm^{-3}} \ B_{\rm MHz}}{\nu_{\rm GHz}^3} \right).
\end{align}   
%%%%%%%%%%%%%%%%%%%%%%%%%%%%%%%%%% 
With $B \approx 3.4$~GHz, $\rm DM = 1775$~pc~cm$^{-3}$, and $\delta {\rm DM} = 0.5$~pc~cm$^{-3}$, we have $t_{\rm chan} \approx 6$~$\mu$s, and $t_{\rm BW} \approx 62$~$\mu$s at 6.1~GHz. Assuming $W_{\rm int}$,~$\tau_{\rm sc} \gg t_{\rm chan},~t_{\rm BW},~t_{\rm samp}$, Equation~\ref{eqn6} can then be reduced to
%%% EQUATION 9
\begin{align}\label{eqn9}
W_{\rm eff} \approx \left(W_{\rm int}^2 + \tau_{\rm sc}^2 \right)^{1/2}.    
\end{align}
%%%%%%%%%%%%%%%%%%%%%%%%%%%%%%%%%% 
For an intrinsically symmetric burst profile, $\tau_{\rm sc} (\nu)$ therefore captures any asymmetry observed in single pulse time series.\\
%%% FIGURE 4
\begin{figure}[t!]
\centering
\includegraphics[width=0.49\textwidth]{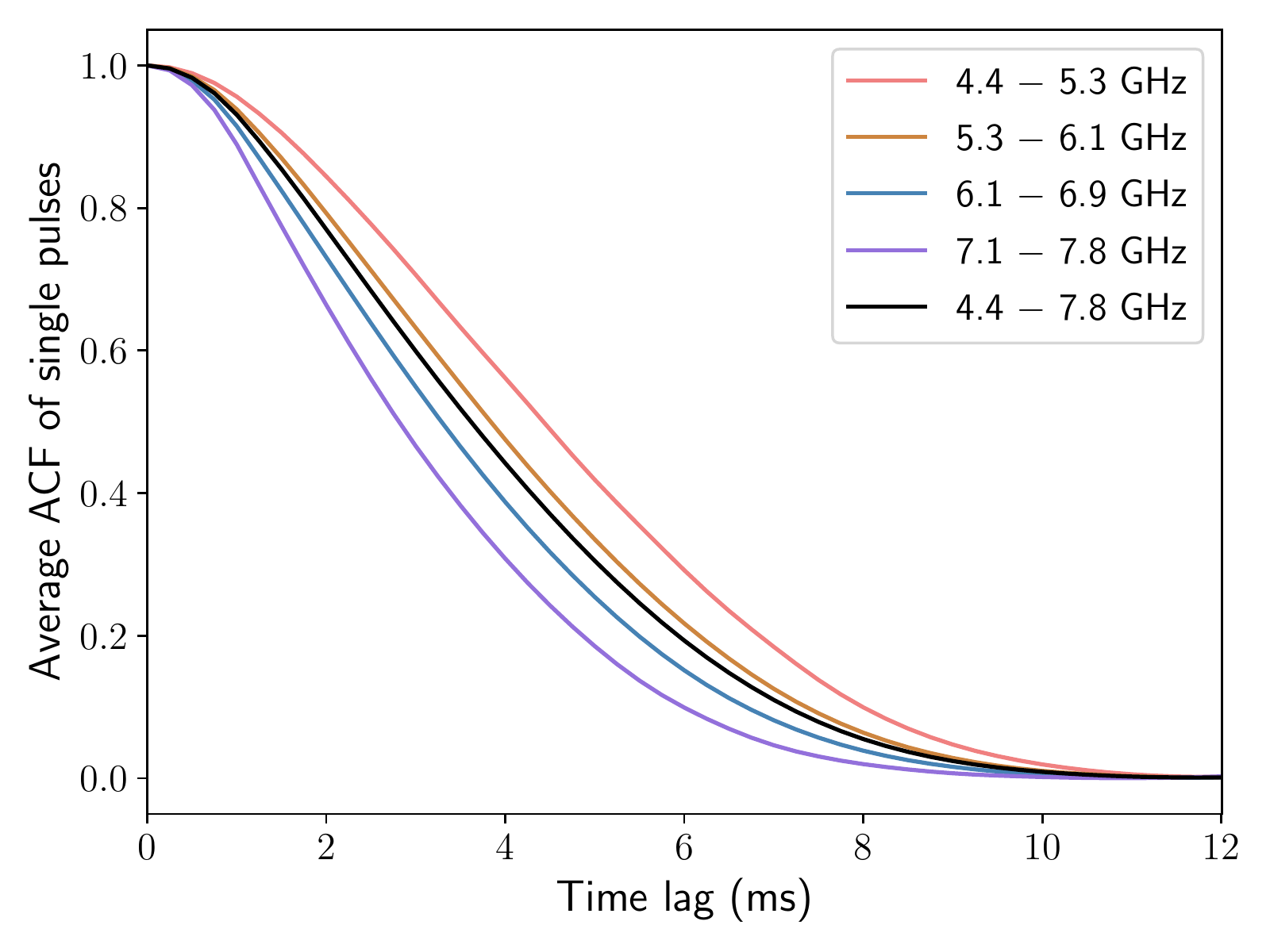}
\caption{Pulse-averaged auto-correlation functions (ACFs) for different radio frequency bands. Plotted ACFs are averages over 2194~pulses detected with (S/N)$_{\rm mf} \geq 8$ in scans 3.1--3.6. We have normalized all ACFs to unit maximum after removal of their noise spikes at zero lag.  \label{fig4}}
\end{figure}
%%%%%%%%%%%%%%%%%%%%%%%%%%%%%%%%%% 

We measured $W_{\rm eff} (\nu)$ through an auto-correlation study of burst time series \citep{Cordes1979,Bartel1980}. To do so, we first split our observing band (excluding flagged channels) into four quadrants based on our instrumental bandpass shape. These quadrants span radio frequency bands 4.4--5.3, 5.3--6.1, 6.1--6.9, and 7.1--7.8~GHz. For each individual quadrant as well as for the entire usable band (4.4--7.8~GHz), we computed their respective frequency-averaged flux density time series, $S(t)$, and normalized these to zero off-pulse mean.\\
%%%% TABLE 3
\begin{deluxetable}{cCcC}
\tablecaption{GC magnetar single pulse widths at different radio frequency bands. Parenthesized numbers represent $1\sigma$ uncertainties on final significant digits. \label{tab3}}
\tablewidth{0.48\textwidth}
\tablehead{
\colhead{Band} & \colhead{$W^{\rm acf}_{\rm eff}$\tablenotemark{a}}  &  \colhead{$W_{\rm skew}$\tablenotemark{b}} & \colhead{$\tau_{\rm sc}$\tablenotemark{c}} \\
\colhead{(GHz)} & \colhead{(ms)} & \colhead{(ms)} & \colhead{(ms)}
}
\startdata
4.4--5.3 & 6.3(5) & \nodata & \nodata \\
5.3--6.1 & 5.4(5) & 3.21(1) & 1.54(1) \\
6.1--6.9 & 4.6(5) & 3.13(2) & 1.28(1) \\
7.1--7.8 & 4.0(5) & 4.5(3), 0.92(2)\tablenotemark{d} & 0.91(4)\\
\hline
4.4--7.8 & 5.1(5) & 3.32(1) & 1.48(1) \\
\enddata
\tablenotetext{a}{Effective pulse FWHM derived from pulse-averaged ACF.}
\tablenotetext{b}{Gaussian pulse FWHM estimated from burst-averaged skewness function.}
\tablenotetext{c}{Scatter-broadening time scale obtained from pulse-averaged skewness function}
\tablenotemark{d}{Intrinsic burst profile modeled as a sum of 2 Gaussians}
\vspace{-8mm}
\end{deluxetable}
%%%%%%%%%%%%%%%%%%%%%%%%%%%%%%%%%%

Let $t_{\rm peak}$ denote the peak time of a single pulse in the 4.4--7.8~GHz band-averaged time series. Considering time samples within a tight window of $\pm 6$~ms around $t_{\rm peak}$, we evaluated the autocorrelation function (ACF) of $S(t)$ for every burst in each quadrant using the numerical version of
%%% EQUATION 10
\begin{align}\label{eqn10}
a(\tau) &= \left< S(t)S(t+\tau) \right> \nonumber \\
&= \int \limits_{t_{\rm peak}~ - ~6~{\rm ms}}^{t_{\rm peak}~ + ~6~{\rm ms}} \text{d}t~S(t)~S(t+\tau).    
\end{align}
%%%%%%%%%%%%%%%%%%%%%%%%%%%%%%%%%% 
Incorporating $N_{\rm pulses} = 2194$~bursts with (S/N)$_{\rm mf} \geq 8$ in scans 3.1--3.6, we next calculated the average ACF of single pulses as
%%% EQUATION 11
\begin{align}\label{eqn11}
A(\tau) = \frac{1}{N_{\rm pulses}}\sum \limits_{n=1}^{N_{\rm pulses}} a_n (\tau).
\end{align}
%%%%%%%%%%%%%%%%%%%%%%%%%%%%%%%%%%
We then cleaned $A(\tau)$ of its noise spike at zero lag due to $t_{R}$. Finally, assuming Gaussian burst shapes\footnote{For a Gaussian of FWHM $W$, the FWHM of its ACF is $\sqrt{2}W$.}, we derived $W^{\rm acf}_{\rm eff}$ at different frequency bands from the full-widths-at-half-maxima (FWHMs) of their respective noise-corrected $A(\tau)$. \\

Figure~\ref{fig4} shows noise-corrected $A(\tau)$ (normalized to unit maximum) for various frequency bands in our study. From Table~\ref{tab3}, we note a growing $W^{\rm acf}_{\rm eff}$ with decreasing $\nu$, a trend consistent with expectations of scattering from multi-path wave propagation through the ISM.

\subsubsection{Single Pulse Asymmetry}\label{sec:skewness}
Having measured $W^{\rm acf}_{\rm eff}$, we estimated $\tau_{\rm sc} (\nu)$ by exploiting the temporal asymmetry of GC magnetar single pulses. To do so, we defined the skewness function \citep{Weisskopf1978,Stinebring1981} of a burst time series $S(t)$ as
%%% EQUATION 12
\begin{align}\label{eqn12}
\kappa(\tau) = \frac{\left<S(t)S^2(t+\tau) \right> - \left<S(t+\tau)S^2(t) \right>}{\left< S^3(t)\right>}.
\end{align}
%%%%%%%%%%%%%%%%%%%%%%%%%%%%%%%%%% 
%%% FIGURE 5
\begin{figure}[t!]
\centering
\includegraphics[width=0.49\textwidth]{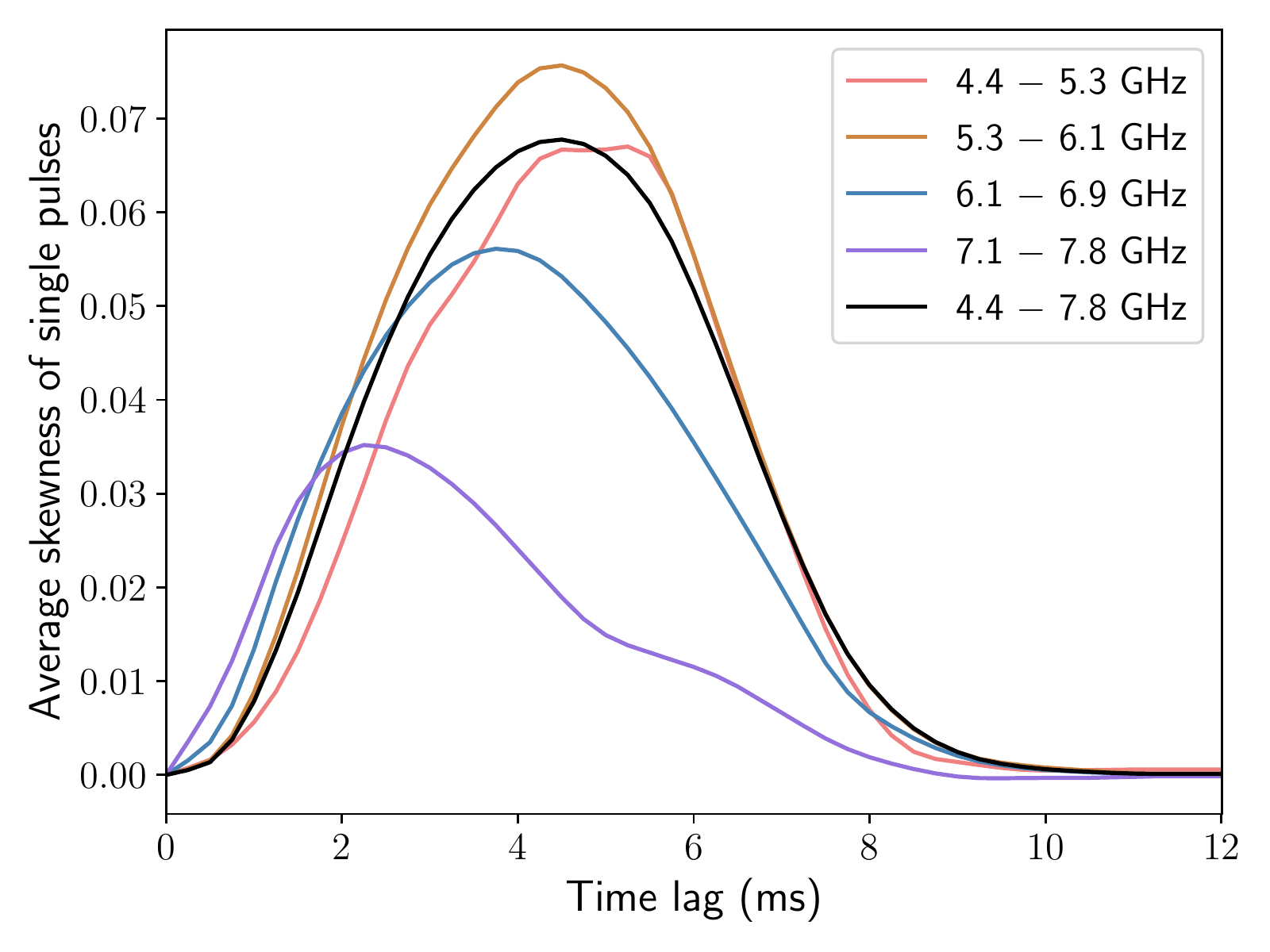}
\caption{Pulse-averaged skewness functions for different radio frequency bands. Plotted curves are weighted averages over 969~pulses detected with (S/N)$_{\rm mf} \geq 15$ in scans 3.1--3.6. \label{fig5}}
\end{figure}
%%%%%%%%%%%%%%%%%%%%%%%%%%%%%%%%%% 
For time-symmetric $S(t)$, $\kappa (\tau) = 0$ by definition. Following Equation~\ref{eqn10}, we again restricted all time integrations in $\kappa(\tau)$ to $\left [t_{\rm peak} - 6~{\rm ms},~t_{\rm peak} + 6~{\rm ms} \right]$ for every burst. However, unlike $a(\tau)$, $\kappa (\tau)$ is insensitive to frequency-averaged burst amplitudes. Therefore, incorporating (S/N)$_{\rm ts}$, the burst S/N in a band-averaged time series, we constructed a weighted pulse-averaged skewness function as follows.
%%% EQUATION 13
\begin{align}\label{eqn13}
K(\tau) = \frac{1}{N_{\rm pulses}} \left( \frac{\sum \limits_{n=1}^{N_{\rm pulses}} {\rm (S/N)}^2_{{\rm ts}, n}~\kappa_n (\tau)}{\sum \limits_{n=1}^{N_{\rm pulses}} {\rm (S/N)}^2_{{\rm ts}, n}} \right).
\end{align}
%%%%%%%%%%%%%%%%%%%%%%%%%%%%%%%%%% 
To further mitigate against noisy contributions to $K(\tau)$ from numerous weak single pulses, we restricted the summations in Equation~\ref{eqn13} to 969~bursts with (S/N)$_{\rm mf} \geq 15$ in scans 3.1--3.6. \\
%%% FIGURE 6
\begin{figure*}[ht!]
\centering
\includegraphics[width=0.98\textwidth]{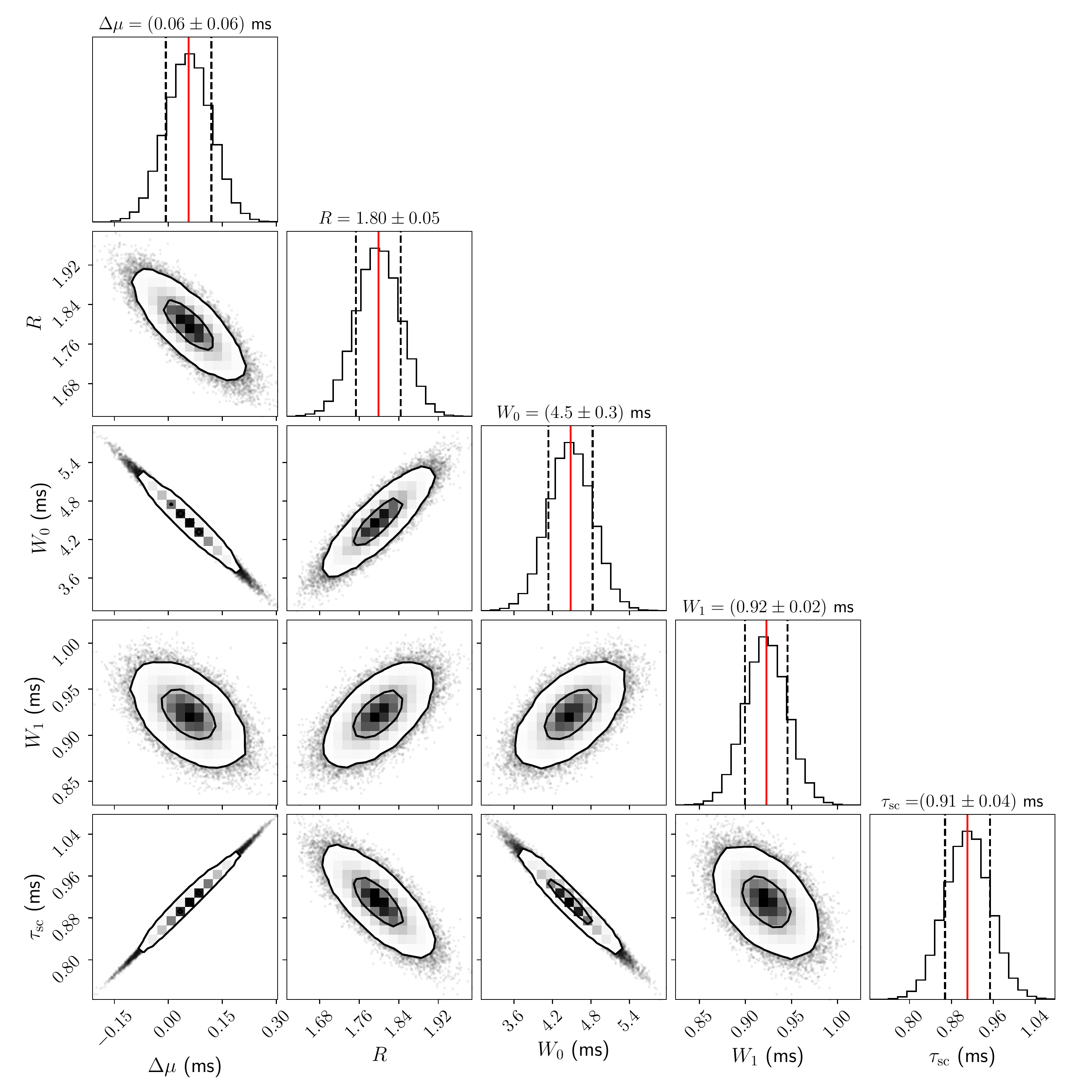}
\caption{2D posterior probability distribution functions (PDFs) of model parameters fit to the pulse-averaged skewness function for the quadrant 7.1--7.8~GHz. In the off-diagonal subplots, the inner and outer black solid contours enclose 50$\%$ and $95\%$ confidence regions, respectively. Diagonal subplots show marginalized 1D PDFs of fit parameters. Red vertical solid lines denote mean parameter values, whereas black vertical dashed lines indicate $\pm 1\sigma$ deviations from their respective means. The fitted model is the skewness distribution of a sum of two Gaussians (numbered 0 and 1) convolved with a pulse broadening function of time scale $\tau_{\rm sc}$. Let $\mu_i$, $W_i$, and $A_i$ denote, respectively, the mean, the FWHM, and the peak amplitude of Gaussian $i$. The fitted parameters are then $\Delta \mu = \mu_1 - \mu_0$, $R = A_1/A_0$, $W_0$, $W_1$, and $\tau_{\rm sc}$. \label{fig6}}
\end{figure*}
%%%%%%%%%%%%%%%%%%%%%%%%%%%%%%%%%% 

Figure~\ref{fig5} shows the ensuing $K(\tau)$ for various frequency bands in our analysis. We interpret $K(\tau)$ in the context of a standard thin screen scattering model \citep{Rickett1977}, described in Appendix~\ref{sec:thinscreen}. As part of our modeling, we treat scattered bursts as native Gaussian pulses convolved with a pulse broadening function (PBF). We further assume that the PBF takes the form of a truncated, one-sided decaying exponential function with time scale $\tau_{\rm sc}(\nu)$. \\

Figure~\ref{figA1} illustrates skewness distributions of scattered pulses for various values of the Gaussian FWHM and $\tau_{\rm sc}$. Comparing Figure~\ref{fig5} with the right panel of Figure~\ref{figA1}, we note that shape of $K(\tau)$ in the frequency bands 5.3--6.1, 6.1--6.9, and 4.4--7.8~GHz qualitatively agrees with that of the skewness function of a scattered pulse with intrinsic Gaussian profile. Fitting our model skewness profile to $K(\tau)$ for these frequency bands, we thus obtained the Gaussian burst FWHM and $\tau_{\rm sc}$ estimates listed in Table~\ref{tab3}, with weak mutual correlations at each band. \\

For scatter-broadened pulses, the amplitude of $\kappa (\tau)$ generally increases with decreasing $\nu$. However, $K(\tau)$ does not conform to the above trend in our bottom quadrant 4.4--5.3~GHz. Informed by our definition of $K(\tau)$ in Equation~\ref{eqn13}, we attribute this discrepancy to low average burst (S/N)$_{\rm ts}$ in the bottom quadrant. Hence, we refrain from extending our simple scattered pulse model to $K(\tau)$ for 4.4--5.3~GHz.\\

Ultimately, we study $K(\tau)$ for our topmost quadrant 7.1--7.8~GHz. Motivated by the anomalous kink in $K(\tau)$ at $\tau \simeq 5$~ms, and the prominence of sub-pulses in dynamic spectra above 7~GHz, we model the intrinsic burst emission as a sum of two Gaussians. Labeling these Gaussians with subscripts $0$ and $1$, let $\mu_i$, $W_i$, and $A_i$, represent respectively, the mean, the FWHM, and the peak amplitude of Gaussian $i$. Since $\kappa (\tau)$ is invariant under translation and scaling of $S(t)$, our burst model now contains five parameters: $\Delta \mu = \mu_1 - \mu_0$, $R=A_1/A_0$, $W_0$, $W_1$, and $\tau_{\rm sc}$. \\

Assuming Gaussian likelihood distributions, and flat, unconstraining priors for all parameters, we performed a non-linear least squares model fit to $K(\tau)$. Figure~\ref{fig6} shows the resulting 2D posterior probability distributions for every pair of parameters. Our model fitting suggests an abundance of bursts comprised of two marginally resolved ($\Delta \mu = 0.06 \pm 0.06$~ms) sub-pulses. Of these sub-pulses, the leading pulse is broad ($W_0 = 4.5 \pm 0.3$~ms) and weak, whereas the trailing pulse is narrow ($W_1 = 0.92 \pm 0.02$~ms) and intense ($R=1.80 \pm 0.05$). Furthermore, all bursts show a scattering tail with $\tau_{\rm sc} \simeq 0.91$~ms at 7.1--7.8~GHz. To within $1\sigma$ uncertainty, our narrowband $\tau_{\rm sc}$ measurements in Table~\ref{tab3} are consistent with the broadband scaling law presented in Equation~\ref{eqn1}. \\

The strong correlations between model parameters in Figure~\ref{fig6} follow from the finite nature of typical burst widths and fluences. For example, at fixed $\Delta \mu$ and $R$, an increase in $\tau_{\rm sc}$ necessitates strongly peaked Gaussian sub-pulses of decreased widths. Alternatively, holding $W_0$, $W_1$, and $R$ constant, a rise in $\tau_{\rm sc}$ can be offset by an accompanying increase in $\Delta \mu$ to preserve the total burst width.

\subsubsection{Single Pulse Fluence}\label{sec:fluence}
\begin{figure}[t!]
\centering
\includegraphics[width=0.49\textwidth]{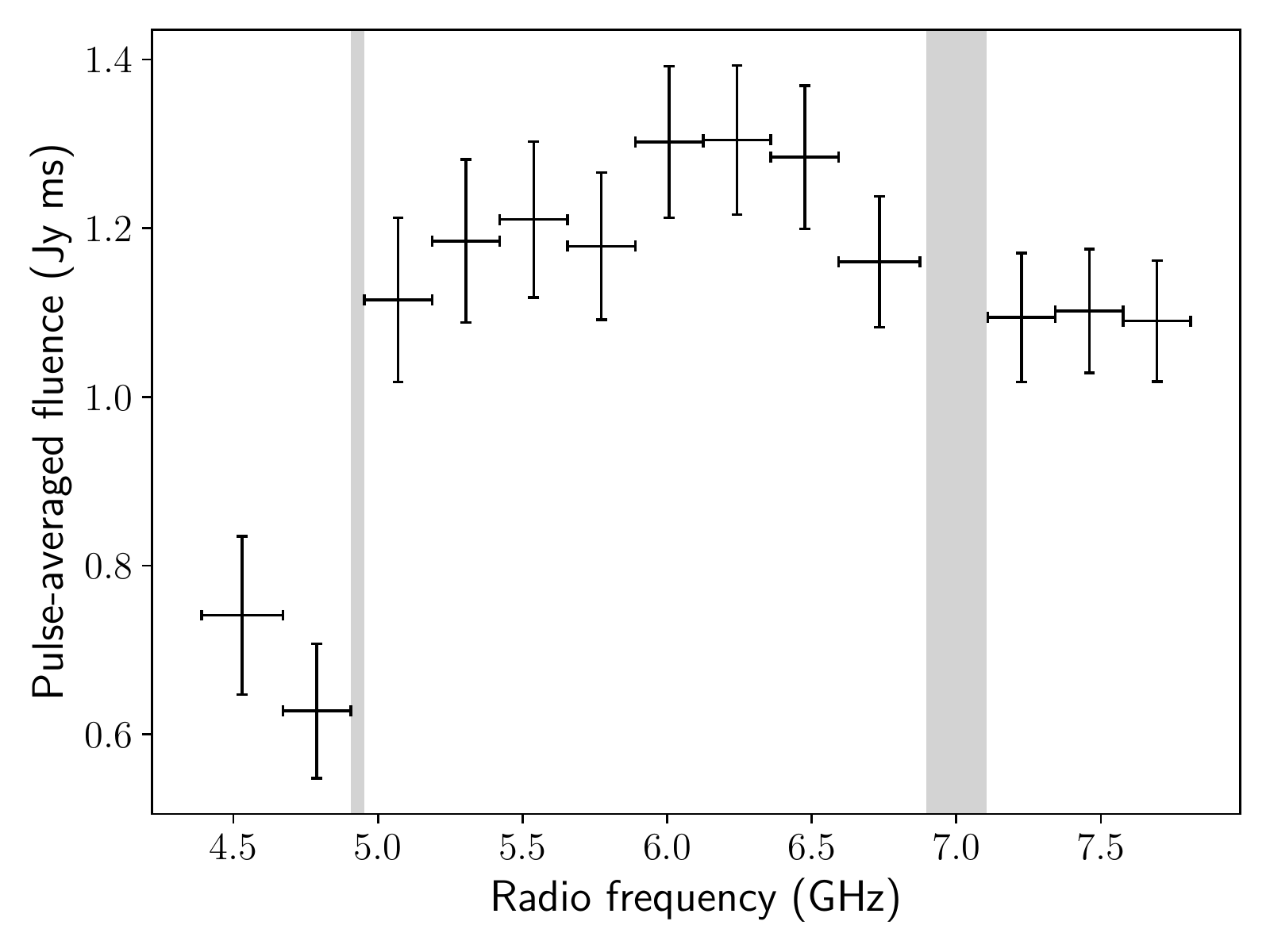}
\caption{Pulse-averaged fluence spectrum of the GC magnetar. Plotted fluences are averages over 2194~pulses with (S/N)$_{\rm mf} \geq 8$ in scans 3.1--3.6. Vertical error bars represent 1$\sigma$ uncertainties on the mean burst fluence. Horizontal error bars indicate subband frequency ranges. Grey vertical bands mark radio frequencies flagged by our RFI detection procedure. \label{fig7}}
\end{figure}

We explored the in-band spectrum of GC magnetar radio pulses using 13 non-overlapping subbands of width, $B_{\rm sub} \approx 234$~MHz, to cover our entire usable band. For each burst in every subband, we computed the subband-averaged burst fluence according to
%%% EQUATION 14
\begin{align}\label{eqn14}
\mathcal{F}_{\rm sub}(\nu) = \frac{1}{B_{\rm sub}} \int \limits_{B_{\rm sub}} \text{d}\nu \int \limits_{t_{\rm peak}~ - ~6~{\rm ms}}^{t_{\rm peak}~ + ~6~{\rm ms}} \text{d}t~S_{\nu} (t).
\end{align}
%%%%%%%%%%%%%%%%%%%%%%%%%%%%%%%%%% 
Averaging $\mathcal{F}_{\rm sub}(\nu)$ over 2194~bursts with (S/N)$_{\rm mf} \geq 8$ from scans 3.1--3.6, we obtained the pulse-averaged fluence spectrum shown in Figure~\ref{fig7}.\\
%%% FIGURE 7
%%%%%%%%%%%%%%%%%%%%%%%%%%%%%%%%%% 

Unlike typical GP \citep{Argyle1972, Popov2007, Karuppusamy2010} and radio magnetar spectra \citep{Torne2015, Lower2021}, our GC magnetar fluence spectrum does not obey a power-law form across our usable bandwidth. In fact, the observed spectrum shows an apparent steep decline below 5~GHz, and remains flat between 5--7.8~GHz to within $2\sigma$ uncertainty. \\

A possible sensitivity issue with our 4.4--5~GHz instrumental bandpass, an unknown complicated $T_{\rm GC}(\nu)$ during our observations, or a spectral break innate to the GC magnetar are all likely causes for the observed emission discontinuity at 5~GHz. Careful distinction between these hypotheses warrants independent high-sensitivity observations of the GC magnetar. \\

Integrating $\mathcal{F}_{\rm sub}(\nu)$ over 5--7.8~GHz, we derive a mean fluence, $\overline{\mathcal{F}}_{6.4} \approx (1.18 \pm 0.03)$~Jy~ms. With 2194~bursts detected in 3~hours (scans 3.1--3.6), $\overline{\mathcal{F}}_{6.4}$ then translates to a 6.4~GHz continuum flux density, $\overline{S}_{6.4} \approx (240 \pm 5)~\mu$Jy for the GC magnetar on MJD~58738.

% SECTION 4: PERIODICITY STUDY
\section{Periodicity Study} \label{sec:periodicity}
%%% FIGURE 8
\begin{figure}[t]
\centering
\includegraphics[width=0.5\textwidth]{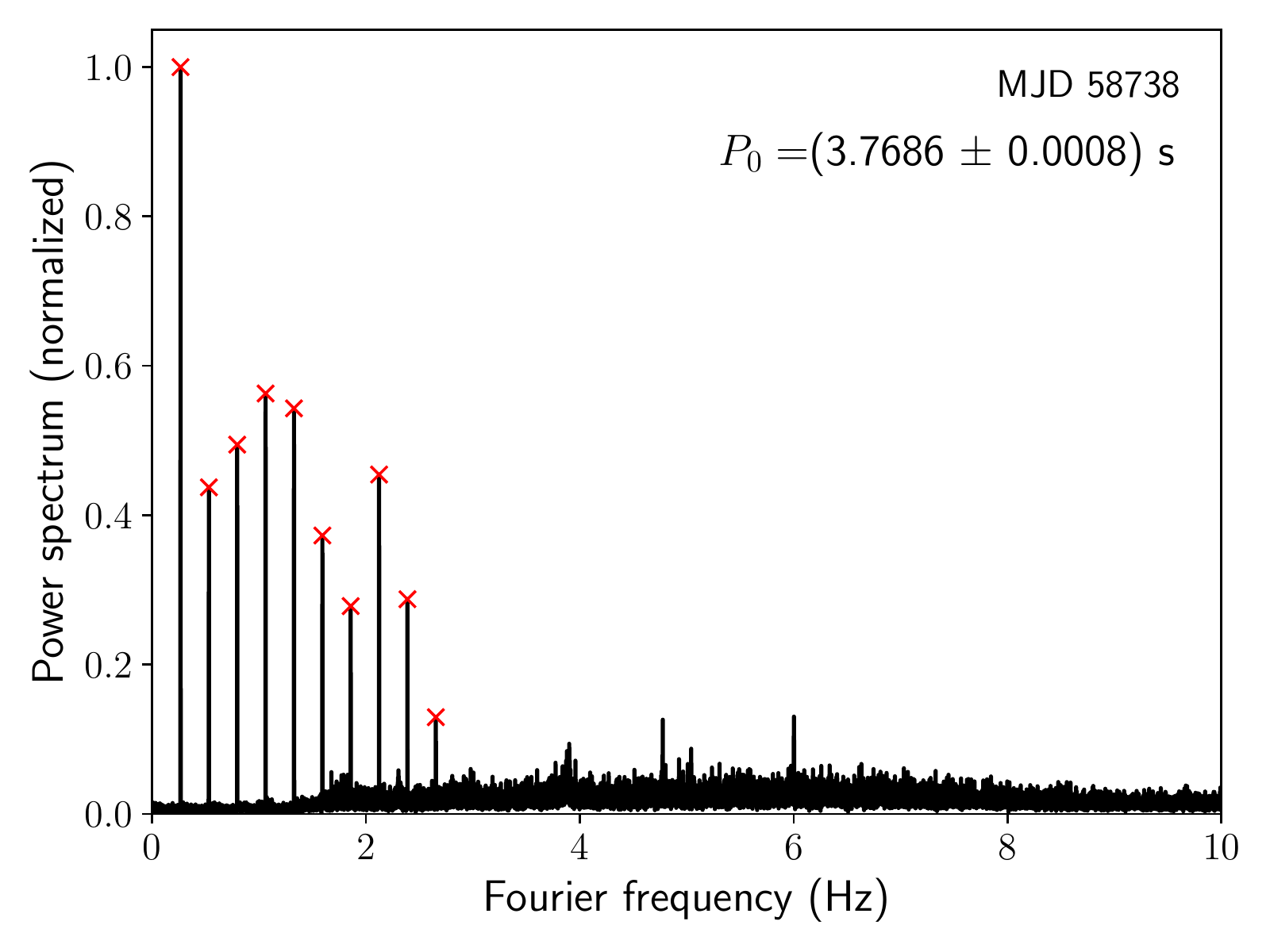}
\caption{Incoherent summation of power spectra of barycentric ${\rm DM} = 1775$~pc~cm$^{-3}$ time series from the eight scans comprising epoch~3 (MJD 58738). The red crosses label the fundamental rotation frequency ($f_0 = 1/P_0$) of the GC magnetar and its harmonics. \label{fig8}}
\end{figure}
%%%%%%%%%%%%%%%%%%%%%%%%%%%%%%%%%% 
Periodicity searches for slow pulsars ($P_0 \geq 1$~s) in long data sets ($\geq$ 10~min.) are often complicated by the presence of low-frequency, power-law noise in Fourier-domain spectra \citep{Ransom2002}. To minimize the deleterious impact of such red noise on our periodicity analyses, we detrended our $\rm{DM} = 1775$~pc~cm$^{-3}$ time series using a running median window of width 0.25~s. We then incorporated barycentric corrections in our detrended time series, and computed their respective Fourier transforms. For epochs consisting of multiple scans, we incoherently summed power spectra from individual scans to increase the S/N of our periodicity detections. \\

Figure~\ref{fig8} shows the incoherently summed power spectrum obtained from eight barycentric $\rm{DM} = 1775$~pc~cm$^{-3}$ time series at epoch~3. We visually identify up to ten harmonics of the fundamental rotation frequency ($f_0 = 1/P_0$) of the GC magnetar at epoch~3. From the highest harmonic of $f_0$ detected at each epoch, we inferred $P_0$ with the greatest possible precision. As noted in Table~\ref{tab2}, our $P_0$ estimates show excellent agreement between epochs, suggesting a consistent GC magnetar timing behavior during our observations. \\
%%% FIGURE 9
\begin{figure}[t]
\centering
\includegraphics[width=0.5\textwidth]{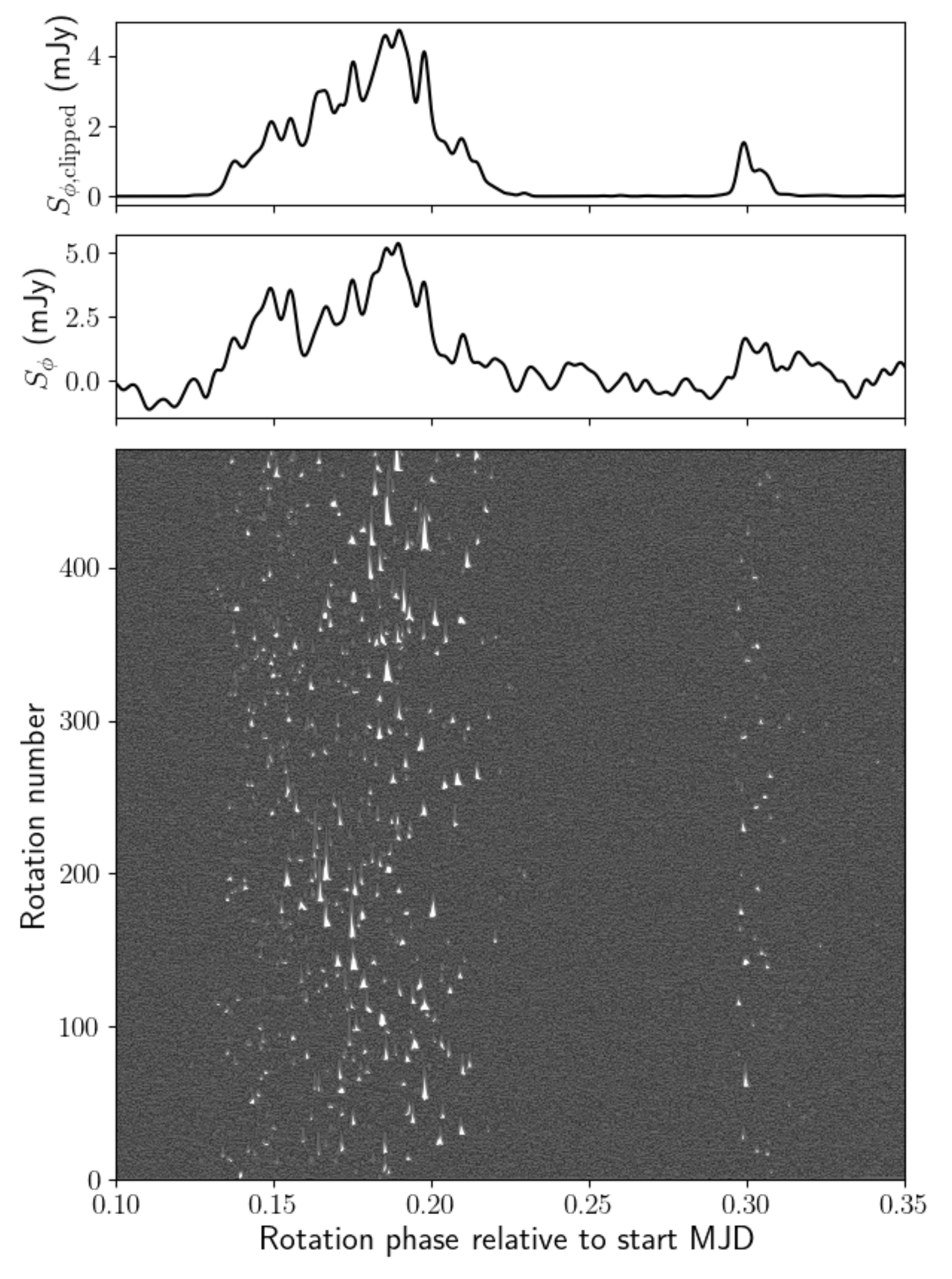}
\caption{4.4--7.8~GHz pulse-averaged profile of the GC magnetar from scan~3.1. The bottom panel shows the rotation-resolved ${\rm DM} = 1775$~pc~cm$^{-3}$ time series, with 4096~phase bins across $P_0 = 3.7686$~s. The middle panel displays the phase-resolved mean flux density profile after smoothing with a Hanning window of length 64~phase bins. GC magnetar single pulses are concentrated in two emission components separated by $\simeq 0.125$~turns ($\approx 471$~ms) in the average profile. To effectively reveal these components, we clipped all data points with ${\rm S/N} \leq 4$ in the dedispersed time series, and replaced these with median values. The top panel shows the average pulse profile obtained by folding the clipped time series. \label{fig9}}
\end{figure}
%%%%%%%%%%%%%%%%%%%%%%%%%%%%%%%%%% 
%%% FIGURE 10
\begin{figure}[t!]
\centering
\includegraphics[width=0.5\textwidth]{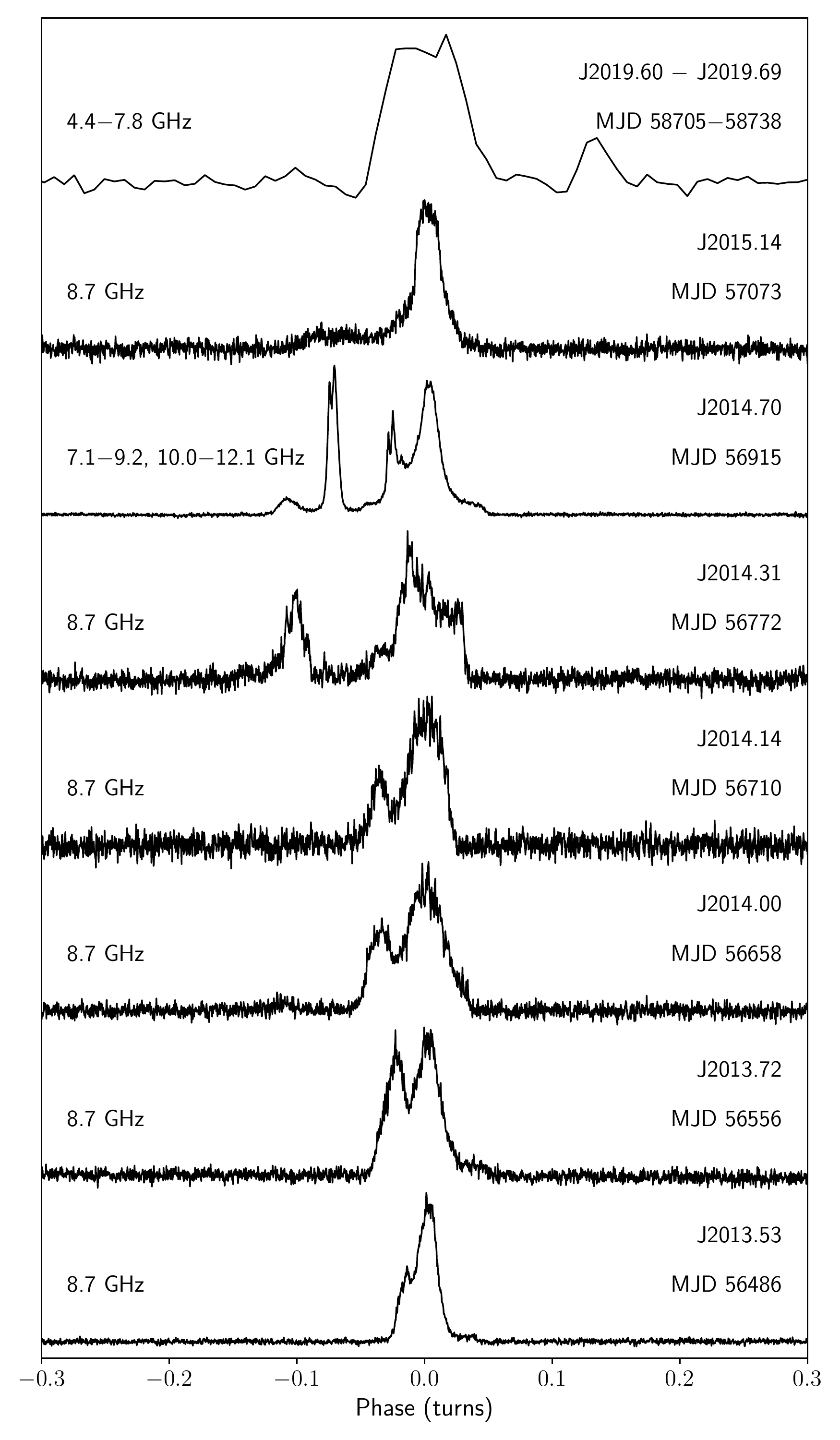}
\caption{Average pulse profile evolution of the GC magnetar over a 6-year period. Profiles for MJDs 56486--57073 have been taken from Figure~2 of \citet{Wharton2019}. Our 4.4--7.8~GHz grand-averaged profile incorporates a net integration time of 6~hours. All 8.7~GHz profiles come from a Very Long Baseline Interferometry (VLBI) campaign \citep{Bower2014,Bower2015} with 6-hour observing sessions, and utilizing 256~MHz of bandwidth. The MJD~56915 profile comes from a 6.5-hour phased VLA observation by \citet{Wharton2019}. We have arbitrarily aligned emission centroids across epochs in the absence of a phase-connected timing solution. \label{fig10}}
\end{figure}
%%%%%%%%%%%%%%%%%%%%%%%%%%%%%%%%%% 

We further examined the emission regularity of the GC magnetar by fitting a timing model to our single pulse arrival times. Accomplishing this exercise with {\tt PINT} \citep{PINT}, we obtained pulse jitter-dominated post-fit residuals ($\sigma_{\rm jitter} \simeq 150$~ms) devoid of timing irregularities (e.g., glitches and anti-glitches). Having thus confirmed the emission periodicity of the GC magnetar, we folded our detrended time series at the $P_0$ values indicated in Table~\ref{tab2}. \\

Figure~\ref{fig9} shows the pulse-averaged profile of the GC magnetar from scan~3.1. Across all scans, GC magnetar average profiles contain two distinct emission components buried within noise. To enhance the detection significance of these components, we combined average pulse profiles from separate scans using the ``shift-and-add'' technique. This methodology involves shifting input profiles by phase lags that maximize their respective cross-correlations with a benchmark profile. Setting the average pulse profile from Figure~\ref{fig9} as a reference, we constructed a grand-averaged profile incorporating data from all scans. \\

Figure~\ref{fig10} compares our 4.4--7.8~GHz grand-averaged profile against 7--12~GHz profiles from MJDs 56486--57073 \citep{Bower2014, Bower2015, Wharton2019}. In the absence of a phase-connected timing solution, we have arbitrarily aligned emission centroids across epochs. The pulse morphology of the GC magnetar continues to gradually evolve over weeks to months, suggestive of a dynamic magnetosphere powering its radio emission.\\
%%% FIGURE 11
\begin{figure}[t!]
\centering
\includegraphics[width=0.5\textwidth]{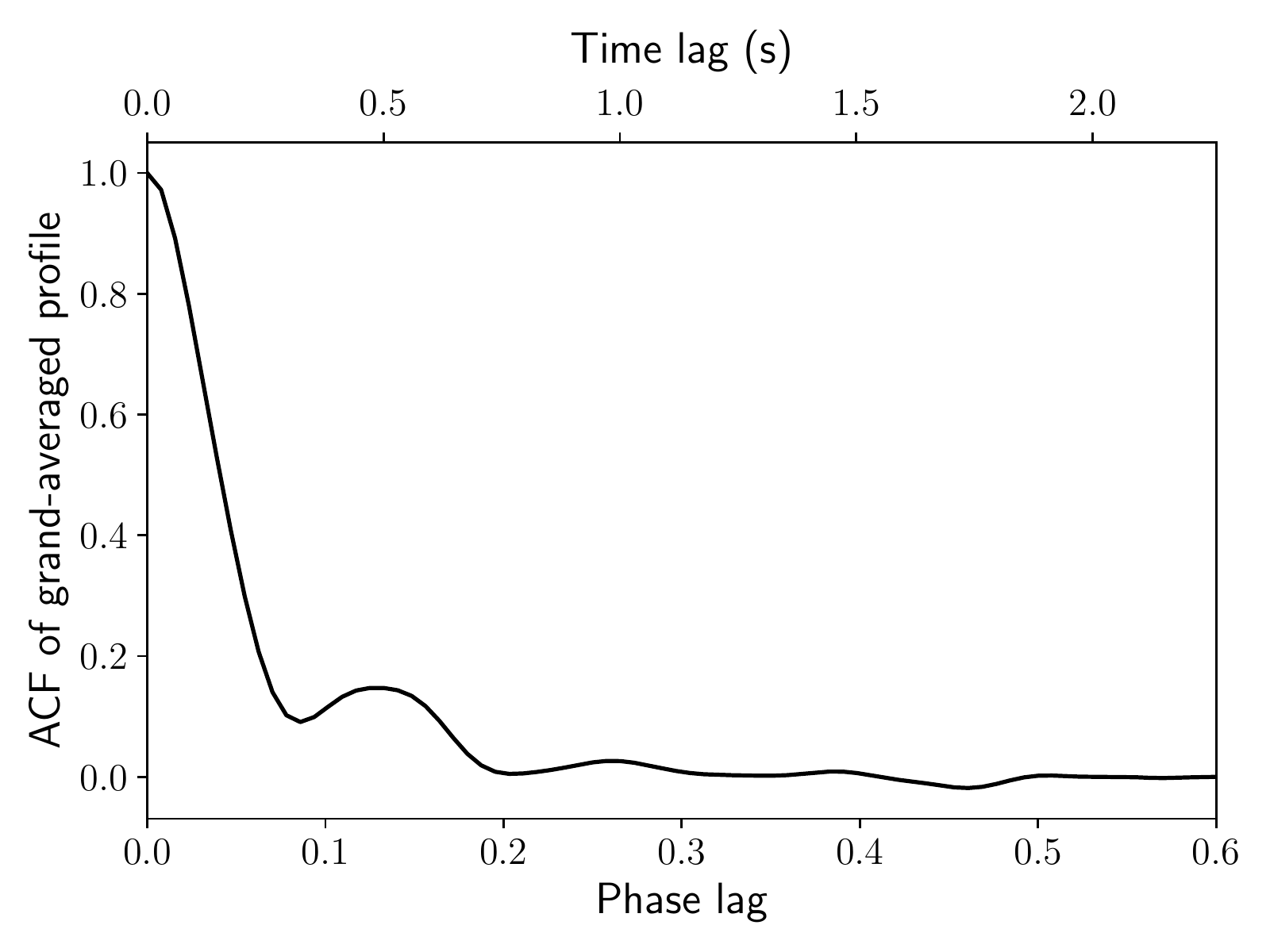}
\caption{ACF of the 4.4--7.8~GHz grand-averaged profile shown in Figure~\ref{fig10}. \label{fig11}}
\end{figure}
%%%%%%%%%%%%%%%%%%%%%%%%%%%%%%%%%% 

Analogous to our single pulse analysis, we characterized sub-structure within our grand-averaged profile through its ACF. Figure~\ref{fig11} shows the normalized ACF of the grand-averaged profile after removal of its noise spike at zero lag. We confirm the existence of two emission components in the average profile from a significant bump in the ACF at phase lag, $\tau_{\phi} \approx 0.125$. For $P_0 \approx 3.7686$~s, this $\tau_{\phi}$ translates to an average component separation of $\approx 471$~ms. \\

Assuming Gaussian component shapes, the FWHM of the primary ACF lobe peaked at $\tau_{\phi}=0$ implies a temporal width, $W_{\rm broad}\approx 220$~ms for the broad emission component in Figure~\ref{fig10}. In comparison, the typical single pulse width inferred from Figure~\ref{fig4} is $\simeq W_{\rm broad}/40$, indicating significant pulse jitter in average profiles. For the narrow emission component, we deduce $W_{\rm narrow}\approx 140$~ms directly from Figure~\ref{fig10}. Finally, we report the statistical consistency of burst properties (widths, asymmetry, and fluence) between the two emission components in the average pulse profile.

% SECTION 5: SUMMARY AND DISCUSSION
\section{Summary and Discussion} \label{sec:disc}
We have conducted a comprehensive study of the 4.4--7.8~GHz emission from the GC magnetar PSR~J1745$-$2900. Using the GBT, we monitored the GC magnetar for a total of 6~hours distributed across MJDs 58705, 58735, and 58738. During our observations, the GC magnetar emitted a flat fluence spectrum over 5--7.8~GHz to within $2\sigma$ uncertainty. Averaging the pulse fluence over 2194~bursts detected in 3~hours, we estimate a 6.4~GHz continuum flux density, $\overline{S}_{6.4} \approx (240 \pm 5)$~$\mu$Jy for the GC magnetar on MJD~58738. \\
%%% FIGURE 12
\begin{figure}[t!]
\centering
\includegraphics[width=0.5\textwidth]{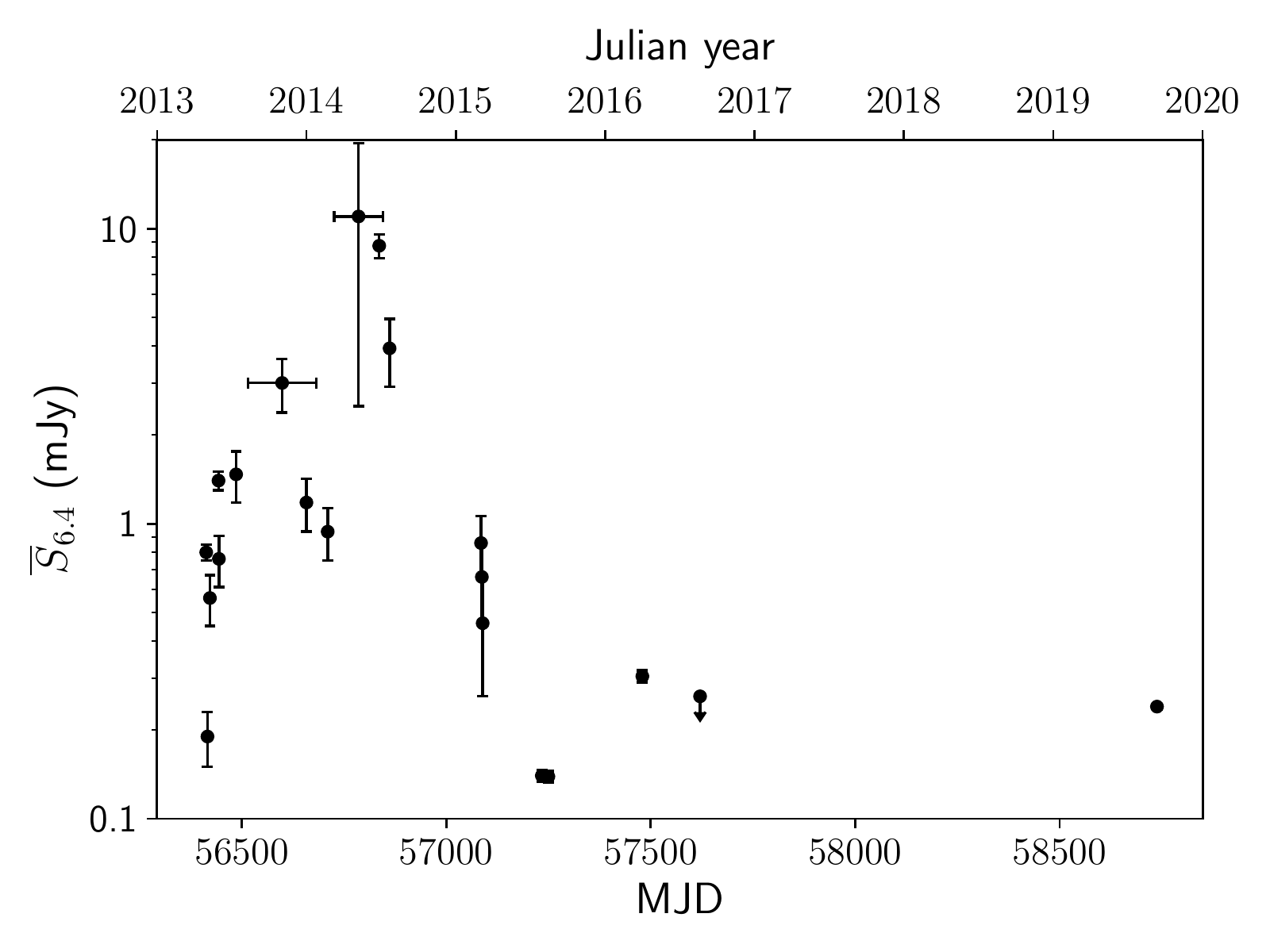}
\caption{Radio variability of the GC magnetar tracked via its 6.4~GHz continuum flux density ($\overline{S}_{6.4}$). For multi-frequency observing campaigns \citep{Eatough2013, Shannon2013, Torne2015, Torne2017, Pearlman2018} that covered 4--8~GHz, we computed $\overline{S}_{6.4}$ through power-law interpolation with spectral indices measured at their respective epochs. For narrow-band 7--9~GHz observations \citep{Bower2015,Lynch2015,Yan2015}, we assumed a flat spectrum to derive $\overline{S}_{6.4}$. \label{fig12}}
\end{figure}
%%%%%%%%%%%%%%%%%%%%%%%%%%%%%%%%%% 

Figure~\ref{fig12} uses $\overline{S}_{6.4}$ as a proxy to trace the radio variability of the GC magnetar. Following a prolonged outburst phase between 2013--2015.5, the GC magnetar has been slowly weakening since 2016. This behavior is consistent with its gradual progress towards X-ray quiescence as reported by the Chandra X-ray observatory \citep{Rea2021}. Table~1 of \citet{Wharton2012} summarizes detection upper limits of GC pulsar searches completed prior to the GC magnetar discovery \citep{Eatough2013} in 2013. The deepest early searches for GC pulsars at 4--8~GHz reach sensitivity thresholds of 17--30~$\mu$Jy, implying that the GC magnetar, as of 2020, is yet to settle to its true ``off'' state. \\

During our observations on MJDs 58705--58738, the GC magnetar exhibited a stable average profile containing two distinct components with FWHMs, $W_{\rm broad} \approx 220$~ms and $W_{\rm narrow} \approx 140$~ms, respectively. Within these emission components, single pulses of much narrower width ($W_{\rm eff} \simeq$ 5 ms) jitter around in the average profile. Such spiky burst emission, while uncommon for radio pulsars, is a noted attribute of radio-loud magnetars \citep{Kramer2007, Levin2012, Yan2015}. \\

Raising particular intrigue, a small subset of single pulses in our data comprise of resolved sub-pulses with different spectral indices, but no detectable radio frequency drifts. To characterize burst sub-structure in dynamic spectra, we manually selected ten bright pulses akin to that shown in Figure~\ref{fig3}(d). For each chosen burst, we computed time-averaged spectra for their constituent sub-pulses. In general, leading sub-pulses within wide bursts follow $S_{\nu} \propto \nu^{1.5 \pm 0.3}$, whereas their trailing counterparts obey $S_{\nu} \propto \nu^{-0.7 \pm 0.2}$ over 5--7.8~GHz. \\

According to our skewness modeling of average single pulse emission at 7.1--7.8~GHz, a typical burst in our sample contains two marginally resolved sub-pulses. Of these sub-pulses, the trailing pulse is $\simeq 1.8$ times brighter than its leading companion. The observed flat fluence spectrum between 5--7.8~GHz can then be rationalized as a confluence between two sub-pulses with opposing spectral signatures. Furthermore, with decreasing radio frequency, scattering increasingly extends the tail of the leading sub-pulse into the head of its trailing counterpart, thus erasing burst structure in dynamic spectra. \\

Proximate sub-pulse emission with contrasting spectral indices is unseen in radio pulsars, but is a prominent feature of magnetar radio emission. For example, \citet{Lower2021} identified the emergence of a flat 0.7--4~GHz spectrum in the magnetar Swift~J1818.0$-$1607 through the gradual superposition of a pulsar-like emission component ($S_{\nu} \propto \nu^{-1.9 \pm 0.2}$) with an inverted spectrum component ($S_{\nu} \propto \nu^{0.4 \pm 0.2}$). While repeating FRBs also show sub-bursts with variable spectral indices \citep{Hessels2019, Pleunis2021}, a notable point of difference is their ``sad trombone'' morphology undetected thus far in radio magnetar spectra. Magnetar models for FRBs \citep{Platts2019}\footnote{FRB theory catalog:~\url{frbtheorycat.org}} must hence explicate the empirical spectro-temporal dissimilarities between Galactic magnetar bursts and FRBs. \\

\citet{Rajabi2020} propose that FRBs arise from intrinsic narrow-band emission processes broadened via relativistic motions. In contrast, radio magnetars generally emit native broadband spectra \citep{Camilo2006,Torne2015, Torne2017}. Our apparent abrupt emission decline below 5~GHz in Figure~\ref{fig7} therefore holds great significance in constraining plausible magnetar and FRB emission models. If deemed astrophysical through independent observations, this spectral turnover could provide clues for unifying Galactic magnetars and FRBs. \\

Finally, we report the statistical consistency of our burst DM and pulse broadening measurements with past estimates ($\rm DM \simeq$~1760--1780~pc~cm$^{-3}$, $\tau_{\rm sc}(\nu) \simeq 1.3$~s $\nu_{\rm GHz}^{-3.8 \pm 0.2}$).  Investigating the large magneto-ionic variations at the GC, \citet{Desvignes2018} invoked a two-screen model for the line-of-sight towards the GC magnetar. A thin plasma screen at $\sim 0.1$~pc from the GC magnetar accounts for its observed $5\%$ fractional $|\rm{RM}|$ variation between 2013--2017. Meanwhile, a second screen at $\simeq 6$~kpc from the GC is responsible for the temporal scatter broadening. \\

Our DM and $\tau_{\rm sc}$ estimates suggest long-term stability of the ISM electron  density spectrum towards the GC. However, continued polarimetric observing of the GC magnetar is essential for examining magnetic field fluctuations in the inner pc of our Galaxy. A detailed understanding of our central ISM not only facilitates modeling of analogous environments towards distant targets such as FRB~121102, but also helps guide future searches for elusive pulsar populations \citep{Wharton2012, Rajwade2017} at the GC. \\

%% IMPORTANT! The old "\acknowledgment" command is now depreciated. It was
\begin{acknowledgments}
AS thanks Robert S.~Wharton for sharing the data required to generate Figure~\ref{fig10}. AS, SC, and JMC acknowledge support from the National Science Foundation (AAG~1815242). Breakthrough Listen is managed by the Breakthrough Initiatives, sponsored by the Breakthrough Prize Foundation. The Green Bank Observatory is a facility of the National Science Foundation, operated under cooperative agreement by Associated Universities, Inc.\\ 
 
This work used the Extreme Science and Engineering Discovery Environment (XSEDE) through allocation PHY200054, which is supported by National Science Foundation grant number ACI$-$1548562. Specifically, it used the Bridges-2 system, which is supported by NSF award number ACI$-$1928147, at the Pittsburgh Supercomputing Center (PSC).
\end{acknowledgments}

%% \facility{} or \facilities{} macros to list the keywords of facilities useD

\vspace{5mm}
\facilities{GBT, XSEDE \citep{XSEDE}.}

\software{Astropy \citep{Astropy2013, Astropy2018},
          corner \citep{corner},
          NumPy \citep{NumPy},
          Matplotlib \citep{Matplotlib},
          {\tt PINT} \citep{PINT},
          {\tt PRESTO} \citep{PRESTO},
          Python~3 (\url{https://www.python.org}), 
          SciPy \citep{SciPy}.
         }

% APPENDIX 
\appendix
\restartappendixnumbering 
% SECTION A1: Thin Screen Scattering Model
\section{Thin Screen Scattering Model}\label{sec:thinscreen}
%%% FIGURE A1
\begin{figure*}[t!]
\centering
\includegraphics[width=0.48\textwidth]{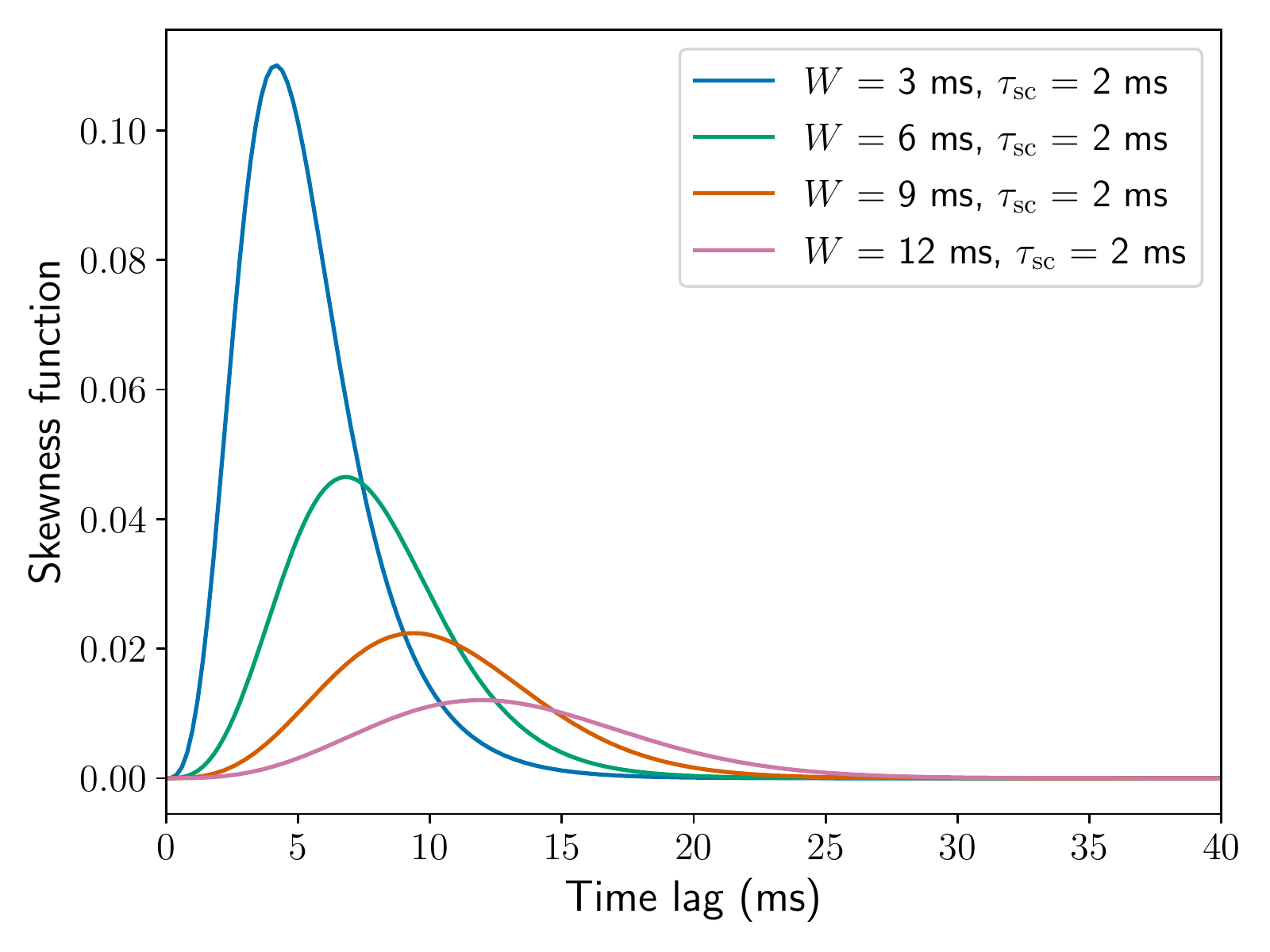}\quad
\includegraphics[width=0.48\textwidth]{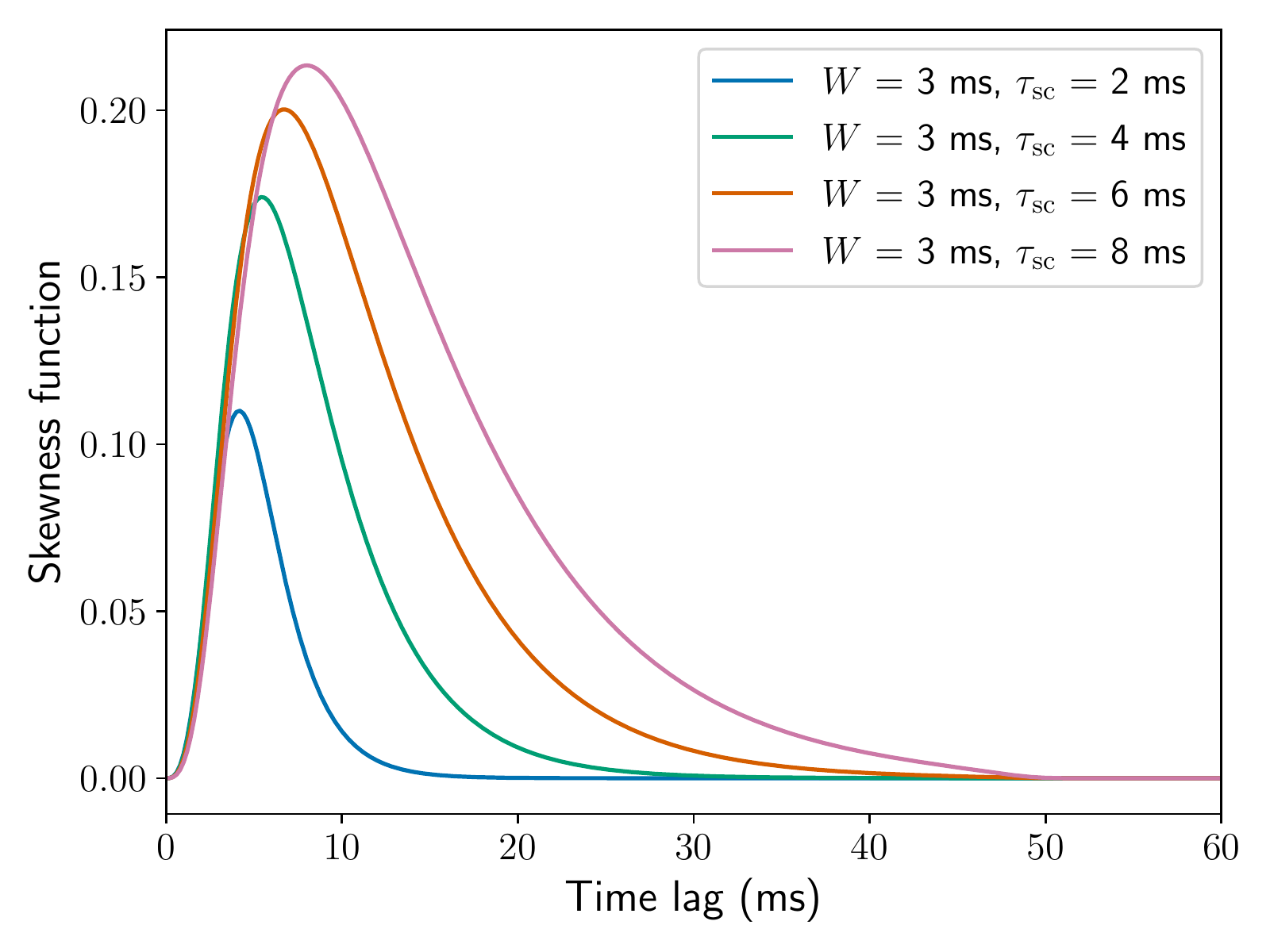}
\caption{Skewness distribution of a scattered pulse with intrinsic Gaussian profile. In the left panel, we fix the pulse broadening time scale, $\tau_{\rm sc} = 2$~ms, and vary the Gaussian FWHM $W$. In the right panel, we vary $\tau_{\rm sc}$, holding $W = 3$~ms constant. For reference, both panels share the same blue curve ($W=3$~ms, $\tau_{\rm sc}=2$~ms). However, note the different axes scales on the two panels. \label{figA1}}
\end{figure*}
%%%%%%%%%%%%%%%%%%%%%%%%%%%%%%%%%% 
Radio pulse broadening via scattering arises from multi-path wave propagation through an inhomogeneous plasma \citep{Rickett1977}. Scattering models for astrophysical environments often invoke thin plasma screens that infinitely extend transverse to the line of sight. Such models encapsulate temporal broadening through a time scale $\tau_{\rm sc}$, involving the source distance and the underlying turbulent electron density spectrum. For electron density fluctuations with a square-law structure function, the pulse broadening function is 
%%% EQUATION A1
\begin{align}\label{eqnA1}
{\rm PBF}(t) &= \frac{1}{\tau_{\rm sc}}e^{-t/\tau_{\rm sc}}H(t).     
\end{align}
%%%%%%%%%%%%%%%%%%%%%%%%%%%%%%%%%% 
Here, $H(t)$ is the Heaviside step function equal to unity for $t \geq 0$, and $0$ otherwise. \\

Consider an astrophysical pulse with native Gaussian profile $G(t)$ of FWHM $W$. Assuming perfect instrumental response and negligible post-detection dispersive smearing, the observed burst profile is given by the convolution,
%%% EQUATION A2
\begin{align}\label{eqnA2}
P_{\rm obs}(t) = G(t) * {\rm PBF}(t).
\end{align}
%%%%%%%%%%%%%%%%%%%%%%%%%%%%%%%%%% 
Equation~\ref{eqn12} defines the skewness function $\kappa (\tau)$, which quantifies the asymmetry of a time series. For a symmetric time series such as $G(t)$, $\kappa(\tau) = 0$. We explore skewness distributions of $P_{\rm obs}(t)$ for different $W$ and $\tau_{\rm sc}$ in Figure~\ref{figA1}. As expected, the peak amplitude of $\kappa$ shows a positive correlation with $\tau_{\rm sc}$, but a negative correlation with $W$. In addition, for peak time lag $\tau_{\rm peak}$, $W$ controls the slope of $\kappa$ at $\tau < \tau_{\rm peak}$, whereas $\tau_{\rm sc}$ regulates the same at $\tau > \tau_{\rm peak}$.
%%%%%%%%%%%%%%%%%%%%%%%%%%%%%%%%%% 
 
\bibliography{references}{}
\bibliographystyle{aasjournal}

\end{document}